\documentclass[journal]{IEEEtran}

\usepackage{xcolor,soul,framed} %,caption
\usepackage[utf8]{inputenc}
\colorlet{shadecolor}{yellow}
\usepackage[pdftex]{graphicx}
\graphicspath{{../pdf/}{../jpeg/}}
\DeclareGraphicsExtensions{.pdf,.jpeg,.png}

\usepackage[cmex10]{amsmath}
\usepackage{array}
\usepackage{mdwmath}
\usepackage{mdwtab}
\usepackage{eqparbox}
\usepackage{url}
\usepackage{multirow}
\usepackage{makecell}
\usepackage{tabularx}
\usepackage{cite}
\usepackage{tikz}
\usepackage{pifont}
\usepackage{algorithmicx}
\usepackage{algorithm}
\usepackage{algpseudocode}
\begin{document}
\bstctlcite{IEEEexample:BSTcontrol}
    \title{QR\text{ï}S: A Preemptive Novel Method for Quishing Detection Through Structural Features of QR}
  \author{Muhammad Wahid Akram, Keshav Sood, \textit{Senior Member, IEEE}, Muneeb Ul Hassan, \textit{Member, IEEE}% <-this % stops a space
% A Preemptive Novel Framework for Quishing Detection Through Structural Features of QR

  \thanks{Muhammad Wahid Akram, Keshav Sood, Muneeb Ul Hassan are with Deakin University, Geelong, VIC 3220, Australia. E-mail: \{s224289198, keshav.sood\}@deakin.edu.au, muneebmh1@gmail.com.}}% <-this % stops a space  

\markboth{}{Akram \MakeLowercase{\textit{et al.}}: }

% ====================================================================
\maketitle

% === ABSTRACT ====================================================================
% =================================================================================
\begin{abstract}
Globally, individuals and organizations employ Quick Response (QR) codes for swift and convenient communication. Leveraging this, cybercriminals embed falsify and misleading information in QR codes to launch various phishing attacks which termed as Quishing. Many former studies have introduced defensive approaches to preclude Quishing such as by classifying the embedded content of QR codes and then label the QR codes accordingly, whereas other studies classify them using visual features (i.e., deep features, histogram density analysis features). However, these approaches mainly rely on black-box techniques which do not clearly provide interpretability and transparency to fully comprehend and reproduce the intrinsic decision process; therefore, having certain obvious limitations includes the approaches' trust, accountability, issues in bias detection, and many more. We proposed QR\text{ï}S, the pioneer method to classify QR codes through the comprehensive structural analysis of a QR code which helps to identify phishing QR codes before-hand. Our classification method is clearly transparent which makes it reproducible, scalable, and easy to comprehend. First, we generated QR codes dataset (i.e. 400,000 samples) using recently published URLs datasets~\cite{prasad2024phiusiil, mendeleyjishnu2024}. Then, unlike black-box models, we developed a simple algorithm to extract 24 structural features from layout patterns present in QR codes. Later, we train the machine learning models on the harvested features and obtained accuracy of up to 83.18\%. To further evaluate the effectiveness of our approach, we perform the comparative analysis of proposed method with relevant contemporary studies. Lastly, for real-world deployment and validation, we developed a mobile app which assures the feasibility of the proposed solution in real-world scenarios which eventually strengthen the applicability of the study.
\end{abstract}

% === KEYWORDS ====================================================================
% =================================================================================
\begin{IEEEkeywords}
QR code, quishing detection, structural feature analysis, phishing, QR\"{\i}S
\end{IEEEkeywords}

\IEEEpeerreviewmaketitle

% === I. INTRODUCTION =============================================================
% =================================================================================
\section{Introduction}

\IEEEPARstart{Q}{R} codes have become the most common and preferred medium of communication in current digital world. Many individuals and organizations such as aviation, hospitality, health, education, banking, and many other sectors, use QR codes to exchange personal contact details, website links, e-ticketing, promotion and branding, etc. In fact, QR codes are evolving into eQR aka executable QR code through embedding a programming language in it~\cite{scanzio2024qr}. This is because, QR codes are easily accessible and mostly can be generated at no cost which makes them economically viable for business activities. Generally, the embedded content of the QR codes is invisible to the target users unless they scan it through a smartphone. Exploiting this, cybercriminals embed phishing content (like URLs) in QR codes and trick victims to scan it with some lucrative approaches~\cite{akram2025exemplifying}. In this case, victims redirected to phishing websites mimicking the legitimate ones or executable URLs instantly download malware in victims' devices~\cite{asiri2024phishingrtds}.\par
Phishing with QR codes is termed as Quishing. Attackers can initiate Quishing attacks in diverse ways such as QRLjacking~\cite{blancaflor2024impact}, barcode-in-barcode attack~\cite{yong2019survey}, and QR code hijacking~\cite{zhou2019invisible}, replacement and manipulation~\cite{bekavac2024qr}. Moreover, in our preliminary work, we have provided detailed information (step-wise) to show how an attacker can use simple open-source tools and LLMs to launch an effective attack~\cite{akram2025exemplifying}. This demonstrated that how a simple two-dimensional matrix of black-white modules in QR code leads user to become a victim through Quishing attacks. To mitigate Quishing, previous research efforts like~\cite{barron2025quashing} utilizes cryptographic signatures to authenticate the embedded payload (i.e. URLs) of the QR codes and~\cite{xue2022screenid} presented SCREENID solution to validate the legitimacy of QR codes through screen specific fingerprints. Although, these efforts are particular reliable in countering tampering or replay attacks, however, these solutions require specialized scanner (because usual built-in and currently available scanners for smartphones are not capable to understand cryptographic signatures and fingerprints), demands specific hardware dependencies, and expect end-users to stay cautious which highlights the infeasibility of these solutions in real-world environments.\par
In addition, few studies~\cite{alaca2023cyber, minocha2024recognition, lourenco2023malicious} presented solutions to visually classify QR codes. Their contributions heavily relies on image-processing and black-box approaches including CNN-based models which compare visual similarity of QR code images and considers deep features and histogram density features to classify them. In another study~\cite{trad2025detecting}, authors proposed a machine learning approach, however, their work relied on pixel-based features which is akin to black-box model techniques (i.e., non-interpretable). With a limited dataset of 9987 samples, authors generated all QR codes of the same version (13), image size (69x69), ECC level (L (Low)), box size (1), and border of QR code (1), respectively. However, these standard settings of their dataset may compromise the robustness and generalizability of their approach in real-world environment. Additionally, by considering each pixel as a feature does not provide any semantic meaning of QR codes' structure like its version, masking pattern, ECC level, etc, which significantly essential for structural classification. Therefore, these approaches are apparently inadequate to provide interpretability and transparency of the intrinsic decision process while classifying QR codes.\par
After our literature review, we note the necessity to develop a method which can be reproducible, scalable, transparent, and real-time effective in diverse  settings irrespective to QR codes version, deployment sector, etc. Considering this, we introduced a novel method named QR\"{\i}S (pronounced as \textit{curious} and short for; \textit{QR: is it Safe?}). In our approach, we extracted structural features of QR codes and classify them accordingly. Also, our Quishing classification method is clearly transparent which makes it reproducible, scalable, and easy to comprehend. Although, we acknowledge that our machine learning (ML) models (XGBoost and Random Forest) relatively obtained average accuracies as compared to black-box techniques. However, it is notable that there is no prior work exists which is able to provide interpretable and transparency details as highlighted in this paper.\par
The main contributions of the paper are listed below:\par
\begin{itemize}
    \item We present a pioneer approach to classify QR codes using the built-in anatomy of QR codes unlike existing black-box model techniques. For this, we extracted 24 features through inspecting the modules of QR codes to fully comprehend the intrinsic decision process while analyzing QR codes as legitimate or phishing. Our technique for extracting these features is generalizable so that it can work on any QR code irrespective of their version.
    \item Our approach is capable to detect QR codes before-hand (without extracting embedded content), which aims to eliminate the necessity of analyzing content (i.e. URLs) and can serve as a preemptive security weapon in context of early detection of phishing attacks.
    \item We provide first-ever datasets of QR code features (i.e., Feat-DataSet-1 and Feat-DataSet-2) containing 400,000 balanced samples. To construct datasets, we generated QR codes from two existing URLs datasets~\cite{prasad2024phiusiil, mendeleyjishnu2024}.
    \item We develop QR\"{\i}S scanner (mobile app) using a cross-platform \textit{Flutter} supporting both android and iOS. Moreover, the experimental results from this app and comparative analysis with existing studies highlight the practical feasibility of proposed framework within the prevailing QR scanners and also in warehouses or retail systems where hundreds of QR codes scanned on daily basis.
\end{itemize}

\textit{Benefit.} In literature, existing solutions are emphasized on analyzing the potentially harmful embedded content of QR codes or they utilize black-box techniques which demands high computational cost~\cite{menghani2023efficient} and makes them unsuitable for real-world deployment in resource-constrained environments (like smartphones). While our QR\"{\i}S solution is a lightweight method which does not require to extract embedded content of QR codes thereby it strengthens users' privacy which is a key factor in protecting end-users. Also, while ensuring generalizability, the proposed solution has the capability to work with any QR code of black and white squares. Also, this solution can be integrated with existing QR scanners. Furthermore, the extracted features are interpretable which can help security specialists to understand the decision making process of the model. Moreover, due to the resource-efficient nature this solution, it can work along within warehouses or retail systems while maintaining the efficiency of work-flow where hundreds of QR codes require to scan on daily basis.\par
\textit{Novelty.} The former relevant studies~\cite{alaca2023cyber, minocha2024recognition, lourenco2023malicious} lack the ability to provide comprehensive details about Quishing classification methods. Although, they achieved high accuracies but further investigation is needed to evaluate whether these studies remain robust with adversarial perturbations or not. In addition, they did not provide any implementation of their devised approaches and in-general their solutions are computationally heavy at the deployment-end. In contrast, we propose a novel approach to mitigate Quishing by extracting structural features of QR code which is able to identify its legitimacy before-hand without extracting embedded content. Moreover, our approach is providing detailed insights of how a QR code is classified as legitimate or phishing. Also, the experimental implementation results and practical feasibility of our solution indicates that the proposed method is lightweight and deployable to resource-constrained devices like smartphones.\par

\section{Literature Review}
We have classified the literature review into three classes; URL-based phishing classification, URL-driven QR classification, QR image-based classification, whereas related work summary is presented in Table~\ref{tab:summary_lit_work}.\par

\begin{table*}[t]
\caption{Summary of Literature Review (LR)}
\centering
\renewcommand{\arraystretch}{1.2}
\begin{tabular}{>{\centering\arraybackslash}m{3.0cm}
                |>{\centering\arraybackslash}m{0.7cm}
                m{4.5cm}                             
                >{\centering\arraybackslash}m{1.3cm}
                >{\centering\arraybackslash}m{1.3cm}
                >{\centering\arraybackslash}m{1.3cm}
                >{\centering\arraybackslash}m{1.3cm}
                >{\centering\arraybackslash}m{1.3cm}}
\hline
\textbf{Type} & \textbf{Ref.} & \textbf{Extracted Features} & \textbf{LR1} & \textbf{LR2} & \textbf{LR3} & \textbf{LR4} & \textbf{LR5} \\
\hline

\multirow{3}{*}[-0ex]{\parbox[c]{3cm}{\centering URL-based phishing classification}}
& \cite{prasad2024phiusiil} & Lexical  & - & - & - & \ding{109} & \ding{109}  \\
\cline{2-3}
 & \cite{aljofey2025bert} & Both lexical and semantic. & - & - & - & \ding{109} & \ding{109} \\
\cline{2-3}
 & \cite{thirumuruganathan2022siraj} & Both lexical and semantic.  & - & - & - & \ding{109} & \ding{109} \\
\hline

\multirow{3}{*}[1.5ex]{\parbox[c]{3cm}{\centering URL-driven QR classification}}
& \cite{marappan2023enhancing} & Lexical & \ding{109} & \ding{109} & \ding{109} & - & \ding{109} \\
& \cite{al2021secure} & Host-based & \ding{109} & \ding{109} & \ding{109} & - & \ding{109} \\
\hline

\multirow{3}{*}[-5.5ex]{\parbox[c]{3cm}{\centering QR image-based classification}}
 & \cite{alaca2023cyber} & 2000 deep features through HHO & \ding{108} & \ding{108} & \ding{109} & \ding{109} & \ding{109} \\
\cline{2-3}
& \cite{minocha2024recognition} & Histogram density analysis for feature extraction & \ding{108} & \ding{108} & \ding{109} & \ding{109} & \ding{109} \\
\cline{2-3}
& \cite{lourenco2023malicious} & Automatic feature selection based on DL & \ding{108} & \ding{108} & \ding{109} & \ding{109} & \ding{109}  \\
\cline{2-3}
& \cite{trad2025detecting} & Pixel-based features & \ding{108} & \ding{108} & \ding{109} & \ding{108} & \ding{109}  \\
\cline{2-3}
& \textbf{Ours} & 24 protocol-level and statistical features of QR code. & \ding{108} & \ding{108} & \ding{108} & \ding{108} & \ding{108}  \\
\hline

\end{tabular}
\label{tab:summary_lit_work}
\vspace{1ex}
\parbox{1\linewidth}{\centering\footnotesize
\textbf{LR1:} \textit{Are they extracting features directly from QR codes?}, \textbf{LR2:} \textit{They do not require to analyze embedded content of QR codes.}, \textbf{LR3:} \textit{Do they provide interpretability details?}, \textbf{LR4:} \textit{Are their solution lightweight?}, \textbf{LR5:} \textit{Have they done any practical implementation of proposed solution?}, (\ding{109}): \textit{No}, (\ding{108}): \textit{Yes.}}
\end{table*}

\subsection{URL-based phishing classification}
In this category, researchers in~\cite{aljofey2025bert, thirumuruganathan2022siraj, prasad2024phiusiil} examined various lexical and domain-based features of URLs and their associated website content for classification. Authors in~\cite{thirumuruganathan2022siraj} presented a unified phishing detection framework based on fine-tune and pre-train paradigm named SIRAJ which is able to work effectively under no labeled or limited dataset. They evaluatef devised approach over diverse phishing sources including phishing and malware URLs, IPs, and malware files. Similarly,~\cite{prasad2024phiusiil} proposed a similarity index and incremental learning based detection framework to identify malicious URLs. In addition to that, few studies tends to look-up blacklists (i.e. Open Phish and Phish Tank) to identify phishing URLs. While being generally effective, these solutions work where URLs are directly visible to individuals and face limitations when it comes to Quishing because URLs are embedded in QR codes.\par

\subsection{URL-driven QR classification}
Attackers leverage QR codes to embed phishing URLs in it which are invisible to individuals until they scan it. Here, few studies~\cite{marappan2023enhancing, al2021secure} emphasize on classifying the QR codes by extracting the embedded content. After thorough analysis, we observed that these studies are pursuing the traditional approaches where phishing URLs are considered as primary attack vectors. They have labeled QR codes as phishing and legitimate based on URL classification. This implies that their contribution do not directly reflect on Quishing (QR code) scenarios where attackers exploiting QR codes as their delivery attack vector. As in Quishing, end-users' often interact visually which makes it challenging to prevent and identify malicious QR codes (containing malicious content).\par

\subsection{QR image-based classification}
In this category, various studies~\cite{alaca2023cyber, minocha2024recognition, lourenco2023malicious} focused on analyzing the visual patterns and features of QR codes to effectively classify them as phishing or legitimate. In~\cite{alaca2023cyber}, authors stated that their approach extracted deep features of QR codes dataset using Harris Hawk Optimization (HHO) algorithm with MobileNetV2 and ShuffleNet. Although authors' work is not based on QR-based phishing attacks, but they provided a black-box model approach for the classification of QR codes generated against 6 different cyber attacks which represents that QR codes can be differentiated visually because of their distinct patterns and features. Another study~\cite{minocha2024recognition}, specifically addressed the challenge of identifying legitimate and phishing QR codes using CNN. Unlike~\cite{alaca2023cyber},~\cite{minocha2024recognition} used histogram density analysis approach to derived various features from QR code images. Similarly, Lourenco et al.~\cite{lourenco2023malicious} also utilized ResNet and VGG19 CNN models to classify QR codes. For feature extraction purpose, they relied on models default capability to automatically determine pertinent features. Likewise, authors in~\cite{trad2025detecting} presented a pixel-based technique akin to black-box to classify QR codes. This is because, their approach does not provide any structural semantic details of QR codes which is crucial for interpretable classification.\par
To sum up, attackers are utilizing diverse tactics for QR based phishing attacks to manipulate individuals to steal their confidential details. Many authors including~\cite{geisler2024hooked, still2024investigating} performed case studies to bolster public awareness about Quishing practices and convey preventive strategies, however, substantial efforts still needed to provoke understanding of diverse Quishing attacks. Moreover, the countermeasures introduced by the former studies might not withstand against Quishing and the further reproducibility is questionable. Therefore, the solution we devised is crucial and timely offers protection to individuals against QR based phishing attacks.

\section{The Fundamentals of QR Codes and Rational to Use Structural Features for Our Study}
To support a clear understanding of our proposed solution ``QR\"{\i}S", this section elaborates the basic fundamentals of QR codes and how they are structurally formed, and further providing insights on QR code modules which can be helpful in classifying QR codes from each other.
\subsection{Anatomy of QR codes}
The two dimensional anatomy of QR codes has total of 40 versions. Except version 1, every QR Code has 9 various regions of their own distinct operations as shown in Fig.~\ref{qr_code_anatomy}. These regions can be grouped into two segments: the function patterns and the encoding segment. The function patterns encompasses finder patterns (positioning pointers), alignment patterns, timing patterns, separators, and quiet zone. While the encoding segment holds format information, version information, data and error correction codes, and remainder bits. Moreover, the description of each region is provided below:

\begin{figure}[htbp]
\begin{center}
\includegraphics[width=\linewidth]{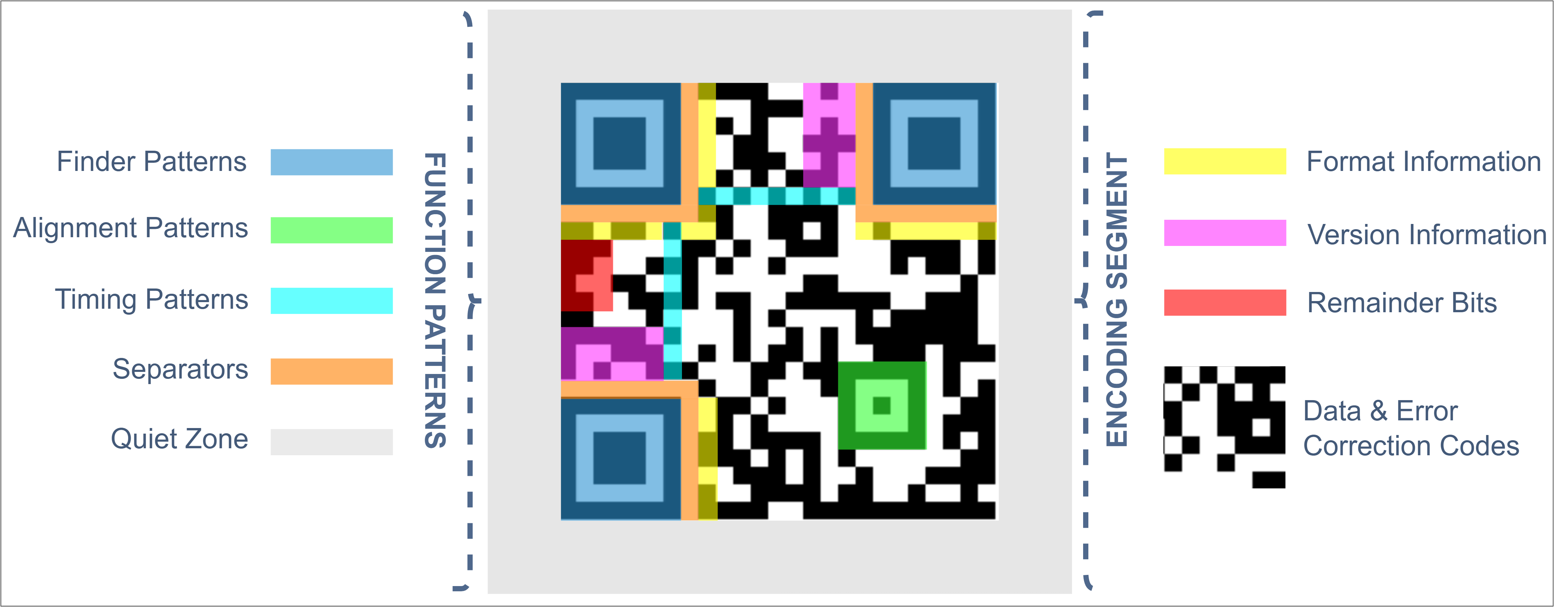}
\caption{An example anatomy of Version 2 QR code. Every QR code consists of two segments that are composed of 9 different regions. These regions represents various information that can be used as features of QR codes.}
\label{qr_code_anatomy}
\end{center}
\end{figure}

\begin{itemize}
    \item \textbf{Finder patterns:} help the scanner to identify the QR Code and its precise positioning. They generally found in corner sides of a QR code with an arrangement of 7x7 modules in black bordered with modules in white which are further bordered with modules in black, except the lower right corner.
    \item \textbf{Alignment patterns:} ensure its contribution in flawless decoding the QR code, even though the adequate distortion occurs in acquired image.
    \item \textbf{Timing patterns:} are visible in an absolute sequence of black and white squares alternatively and facilitates the scanner in evaluating the size of each module.
    \item \textbf{Separators:} comprises of white modules and detaches finder patterns from data modules to enhance their clarity.
    \item \textbf{Quiet zone:} depicts at least 4 modules wide white space margin around a QR code. It also discriminates the code from its surrounding area (among images, text).
    \item \textbf{Format information:} is in 15 bits size reserving details about error correction codes (ECC) level (a feature to retrieve preserved data regardless of designated percentage of damage in QR such as Low (L) 7\%, Medium (M) 15\%, Quartile (Q) 25\%, and High (H) allows 30\%) and the specified masking pattern (a righteous arrangement of modules to refine scan-ability).
    \item \textbf{Version information:} keeps record of the version employed in QR code's formation, which relates to its data volume and size.
    \item \textbf{Data and error correction codes:} contains the actual data encoded in QR code. Both codes have a word length of 8 bits. Moreover, before encoding, the actual data needs to transform into the bit stream.
    \item \textbf{Remainder bits:} (spare bits) utilized only when total bits of both data and error correction are not multiple of 8.
\end{itemize}

\subsection{Understanding of QR Code Modules}
Every QR code is a fusion of black (1's) and white (0's) modules. Modules represents encoding data in a QR code, unlike pixels which are part of display resolution. The modules should not be confused with pixels, in fact, a single module can be a combination of many pixels. These modules are dynamically arranged for QR codes in different regions in all versions which also indicates their data capacity such as version 1 QR code has 21x21 modules, version 2 has 25x25 and every next version increments further with 4 modules.\par

\begin{figure}[htbp]
\begin{center}
\includegraphics[width=\linewidth]{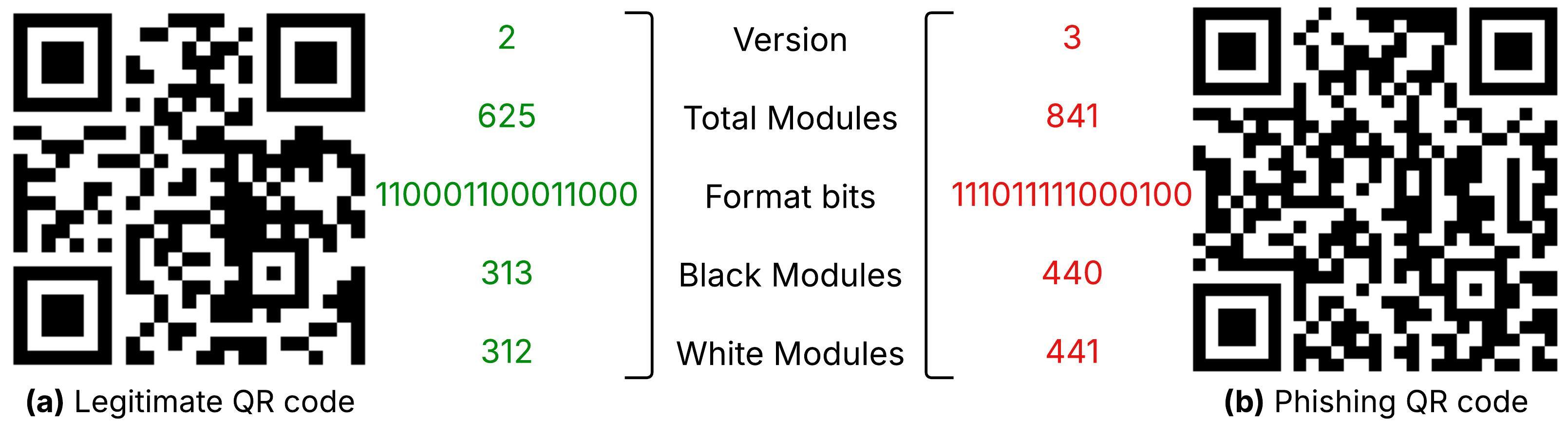}
\caption{A general understanding of differentiation between QR codes. (a) represents a legitimate and (b) represents a phishing QR codes. Both QR codes have clear difference in their in-built structural features.}
\label{comparing_legit_phish_qr}
\end{center}
\end{figure}

In our study, we highlighted the patterns of QR codes as features and further demonstrated that how they can be employed to classify QR codes (i.e. legitimate or phishing). For example, Fig.~\ref{comparing_legit_phish_qr}, (a) represents a legitimate QR code containing valid URL and (b) is a phishing QR code having malicious URL. It can be observed that both QR codes has different version, format information modules (bits), and total number of modules. In addition, after further analysis of considering other statistical features including black into white modules ratio, number of black and white modules, their density, mean density, and more, we established that all of these features can be potentially useful to achieve our objective of classification of QR codes to identify them as legitimate and phishing. It is worth noting that we are not focused on the features of embedded content of QR codes, in contrast, we emphasize on distinguish the legitimacy of QR codes by extracting and analyzing their various structural features before-hand (i.e., before accessing embedded malicious content) after scanning a QR code which has never done before in the existing literature. Furthermore, the comprehensive detail of extracted features and their extraction process is provided in Section IV-B.

\section{Dataset Creation and Features Analysis}
As stated before that we used structural features in our methodology, so data collection was a critical step. For this, we have created our own two datasets since our solution (QR\"{\i}S) is novel and no existing datasets are publicly available that is applicable for it.

\subsection{Dataset Preparation}
\begin{figure}[htbp]
\begin{center}
\includegraphics[width=\linewidth]{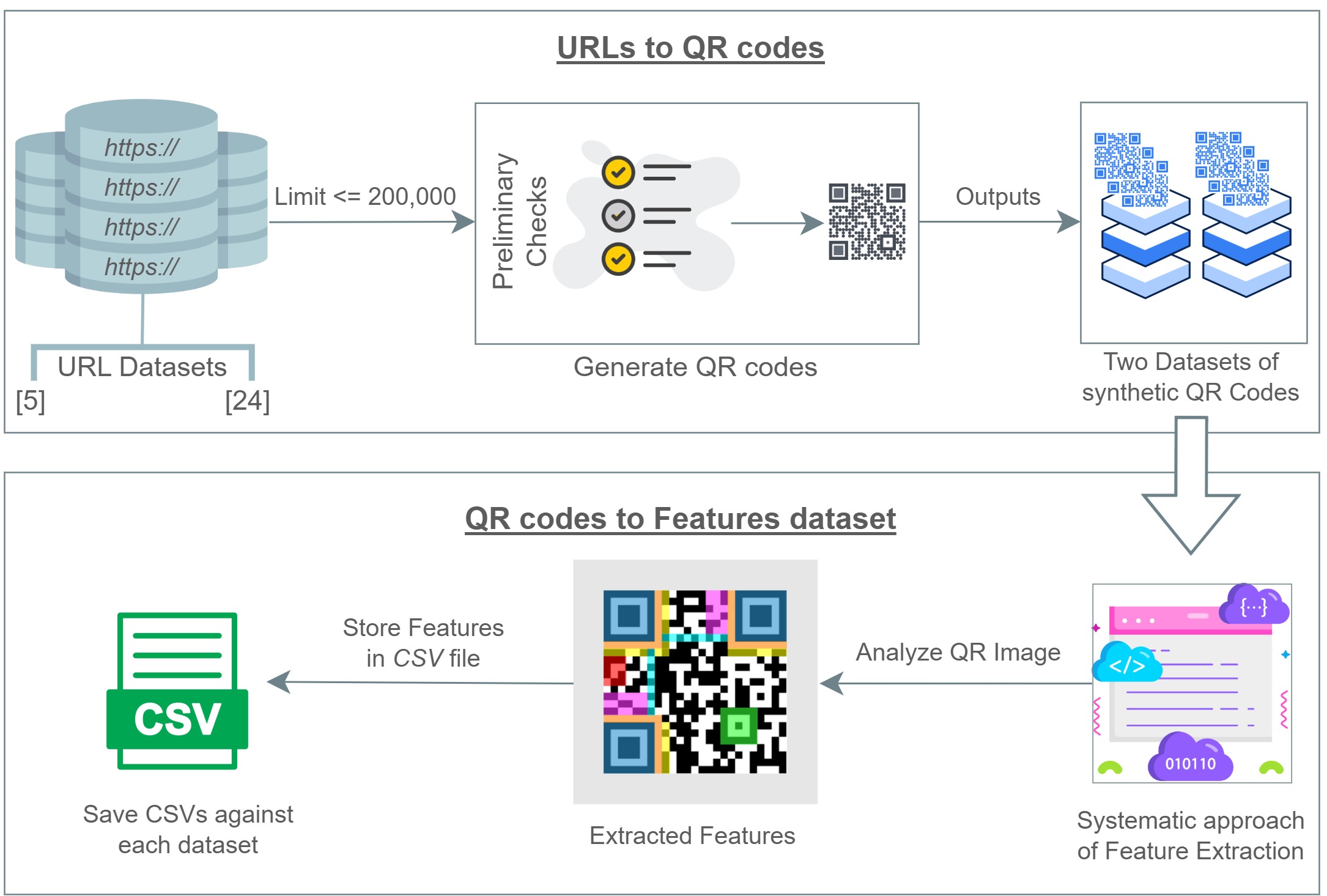}
\caption{An overview of utilizing URLs datasets and converting them in to synthetic QR codes. This process involves multiple preliminary checks to ensure smooth generation of QR codes. Afterwards, these QR codes are fed into a systematic approach which extract protocol-level and statistical features and saving in a CSV file.}
% \vspace{-17pt}}
\label{operational_workflow_of_dataset_creation}
\end{center}
\end{figure}

For this, we thoroughly search in the existing literature and online repositories for an up-to-date dataset of QR codes. We noticed that available QR codes datasets are either outdated~\cite{qr_codes_dataset} or encompassing insufficient samples~\cite{trad2025detecting, minocha2024recognition}. Consequently, we opted to generate two new datasets of synthetic QR codes from recently published and publicly available URLs datasets~\cite{prasad2024phiusiil, mendeleyjishnu2024}. We extracted features from both synthetic QR codes datasets and prepared our features datasets named as Feat-DataSet-1 and Feat-DataSet-2. Both features datasets are different in terms of having varying distribution and patterns of QR codes that can help us to evaluate the effectiveness, robustness, and generalizability of our proposed solution. The distribution of URLs datasets, generated QR codes datasets, and finally created features dataset of QR code are provided in Table~\ref{distribution_of_datasets} and their operational flow is presented in Fig.~\ref{operational_workflow_of_dataset_creation}.\par

\begin{table}[h]
\caption{Distributions of URLs, synthetic QR Codes and Features Datasets}
\centering
\renewcommand{\arraystretch}{1.2}
\begin{tabular}{>{\centering\arraybackslash}m{1.8cm}
                |>{\centering\arraybackslash}m{1cm}
                |>{\centering\arraybackslash}m{1.2cm}
                |>{\centering\arraybackslash}m{1.2cm}
                |>{\centering\arraybackslash}m{1.2cm}}
\hline
\textbf{Dataset Type} & \textbf{Ref.} & \textbf{Legit} & \textbf{Phish} & \textbf{Total} \\
\hline
\hline
\multirow{2}{*}{{{\makecell{URLs}}}} & \cite{prasad2024phiusiil} & 135,850 & 100,945 & 236,795 \\
 & \cite{mendeleyjishnu2024} & 345,738 & 104,438 & 450,176 \\
\hline
QR codes & Ours & 200,000 & 200,000 & 400,000 \\
\hline
Feat-DataSet-1 & \multirow{2}{*}{\makecell{\textbf{Ours} \\ \textbf{(final)}}} & 100,000 & 100,000 & 200,000 \\
Feat-DataSet-2 & & 100,000 & 100,000 & 200,000 \\
\hline
\end{tabular}
\label{distribution_of_datasets}
\end{table}

From Fig.~\ref{operational_workflow_of_dataset_creation}, we provided URLs datasets to a script written in python and also configured it with several preliminary checks to ensure that it reads URLs one-by-one and generate valid synthetic QR codes. For example, QR code natively supports various encoding modes~\cite{kieseberg2010qr} with specific data capacity
to encode data in it 
whereas for a string URL the most appropriate modes are alphanumeric and byte. This implies if a URL is alphanumeric then our script will follow the data capacity limit\footnote{\url{https://www.thonky.com/qr-code-tutorial/character-capacities}} of this mode and similarly, this holds for byte mode. In this way, the script omits URLs which are too long and exceeding the maximum data limit of QR codes. Another check involves choosing ECC level of the generated QR code. We noticed that a URL (either legit or phish) of specific length might be embedded in QR code with different ECC levels. For example, in the dataset of~\cite{prasad2024phiusiil}, a URL\footnote{\url{https://www.drivesmartbc.ca/}} of length 27 characters is possible to encode in a QR code of version 2 with three ECC levels i.e. L, M, Q. In that case, our script dynamically pick ECC level from the possible choices. Moreover, remaining properties (e.g. version, masking pattern) of QR code structure are automatically assigned by the employed library\footnote{\url{https://pypi.org/project/qrcode/}} of QR code. Overall, we generated 200,000 balanced (legitimate and phishing) samples of synthetic QR codes for each of the URLs datasets in~\cite{prasad2024phiusiil} and~\cite{mendeleyjishnu2024}. Furthermore, the next crucial step was to extract features of both synthetic QR codes datasets. We develop a systematic approach in python which accepts, analyzes, outputs, and stores the derived features in a CSV file. Finally we developed two datasets of features (Feat-DataSet-1 and Feat-DataSet-2) each having 200,000 balanced samples and collectively 400,000 samples.\par

\subsection{Features Details and Extraction Approaches}
This section discussed the type of features we extracted and how these features can be identifiable from QR codes. Moreover, these features are also presented in Table~\ref{tab:list_qr_code_features}.\par

\subsubsection{Protocol-level Features}
\begin{figure}[htbp]
\begin{center}
\includegraphics[width=\linewidth]{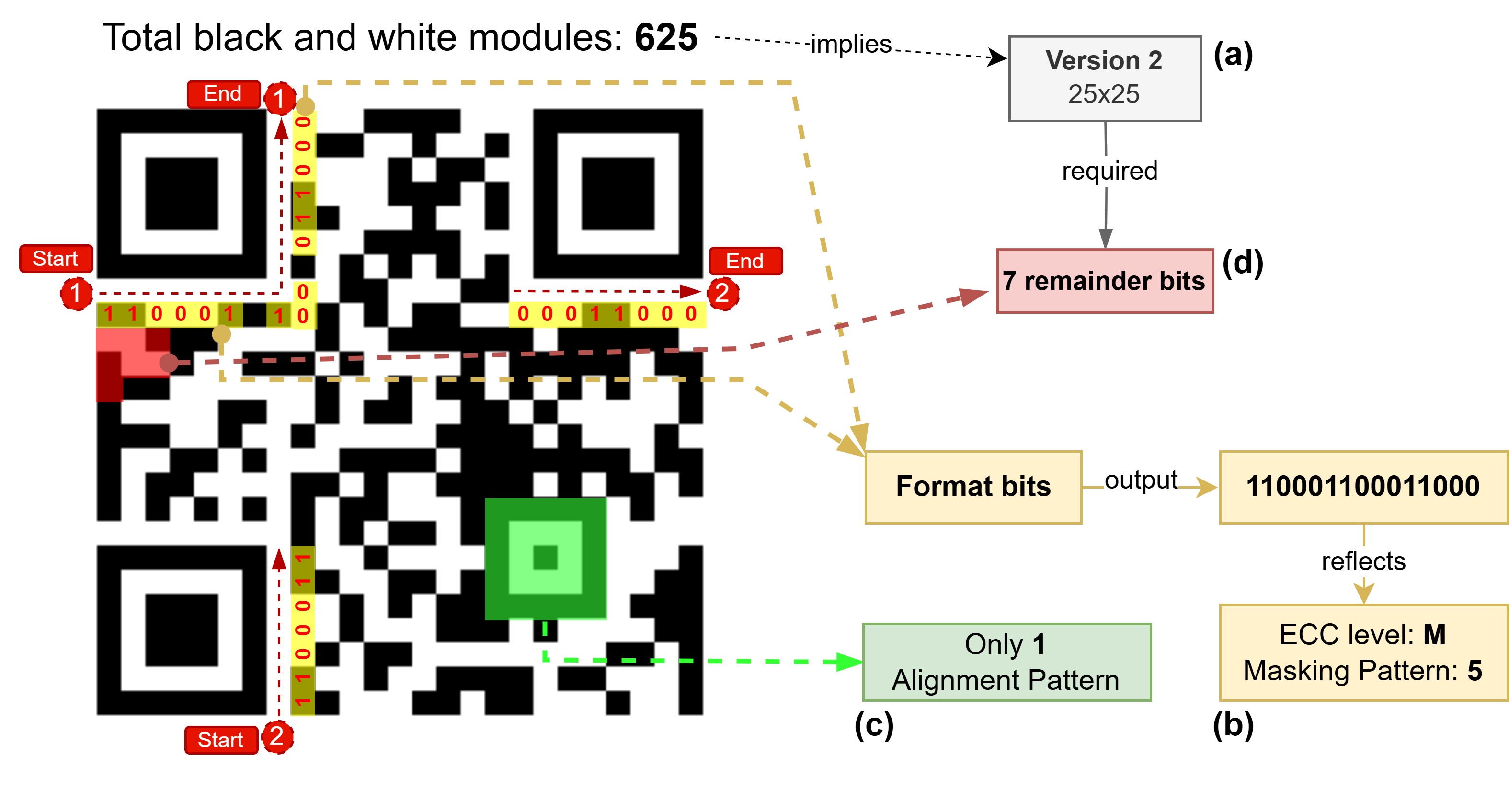}
\caption{Example of capturing protocol-level features of QR codes: (a) Version: calculating total modules, (b) ECC level and masking pattern: found in 15-bits format information, (c) Number of alignment patterns: found in specific positions, (d) Required remainder bits: usually not present in every QR code.}
\label{protocol_level_features}
\end{center}
\end{figure}

These features are related to QR code standard (ISO/IEC 18004) and represents that how a QR code is structurally formed, encoded, and interpreted. In general, except version 1, a QR code has 9 protocol-level features as presented in Fig.~\ref{qr_code_anatomy}. We considered 5 of them because the other 3 features (timing patterns, separators, quiet zone) are static ones meaning there is no dynamic value associated with them. While, in data \& error correction codes feature, the actual content is embedded in it and we are not considering the embedding content of the QR codes in our study. Moreover, the protocol-level features are described below and presented in Fig.~\ref{protocol_level_features}.

\begin{enumerate}
    \item \textbf{Version:} This feature can be derived from any QR code in two ways. In first method which we adapted is to count the number of modules present in the QR code. As previously described, the specific number of modules are standardly pre-defined for all versions of QR codes. For example, from Fig.~\ref{protocol_level_features} (a), a QR code containing 625 modules in total which specifies that it is a version 2 QR code because a version 2 can only have 25x25 modules. For second method, as two purple region of 18 modules are highlighted in Fig.~\ref{qr_code_anatomy}. If we convert this pattern into actual bit string then it will be \textit{000111110010010100}, representing version 7 of the QR code as listed in~\cite{thonky2023}. However, version region is specifically included from version 7 of QR codes, therefore, the second approach is not applicable for every QR code.
    \item \textbf{ECC Level:} The ECC level of a QR code can be found from the format information modules (yellow highlighted region in Fig.~\ref{protocol_level_features} (b)), which is of 15 bits long i.e. \textit{111011111000100}. In fact, format information contains two values ECC level and masking pattern. From~\cite{thonky2023}, it can be confirmed that the captured format bits of QR code has ECC level of `L' with masking pattern as `0'.
    \item \textbf{Masking Pattern:} As mentioned above, the masking pattern can also be extracted from the format information bits which is `0' in case of Fig.~\ref{protocol_level_features} (b). Masking patterns are used to encode almost even distribution of black and white modules in a QR code. This helps scanners in decoding and enhancing the contrast of QR image~\cite{kieseberg2010qr}. Generally, there are 8 standard masking patterns (from 0 to 7).
    \item \textbf{Number of Alignment Patterns:} Except version 1, every QR code has alignment pattern(s) placed in specific positions which helps to fix distortions and support precise decoding when it is tilted, skewed, curved, or printed/displayed on non-planar area~\cite{kieseberg2010qr}. In our case, we counted the number of alignment patterns available in a QR code for the classification purposes. The locations of alignment patterns for every QR code versions are listed in~\cite{thonky2023}.
    For example, in Fig.~\ref{protocol_level_features} (c), QR code is having only 1 alignment pattern.
    \item \textbf{Required Remainder Bits:} For some QR code versions, remainder bits (modules) are added when the final encoded data modules falls short to fulfill the required number of modules in a QR code~\cite{thonky2023, kieseberg2010qr}. We followed the publicly available list of required remainder bits against each version in~\cite{thonky2023}.
    Moreover, the relevant example is presented in Fig.~\ref{protocol_level_features} (d).
\end{enumerate}

\begin{table}[t]
\caption{Extracted Features of QR codes}
\centering
\renewcommand{\arraystretch}{1.2}
\begin{tabular}{>{\centering\arraybackslash}m{1.6cm}
                |m{4.5cm}                           
                |m{1.3cm}}                          
\hline
\textbf{Feature Type} & \textbf{Feature Name} & \textbf{Data Type} \\
\hline
\hline

Protocol-level Features
 & Version, ECC Level, Masking Pattern, Number of Alignment Patterns, Required Remainder Bits & Numerical \\
\hline
\hline

\makecell{Statistical\\Features}
 & No. of White Modules, No. of Black Modules, Black-White Ratio, QR Density, QR Mean Density, QR Standard Density Row-wise, QR Standard Density Column-wise, QR Row Transitions Total, QR Column Transitions Total, QR Entropy, QR Vertical Asymmetry, QR Horizontal Asymmetry, QR Top Left Density, QR Top Right Density, QR Bottom Left Density, QR Bottom Right Density, QR Center Density, QR Row-wise Histogram Peaks, QR Column-wise Histogram Peaks & Numerical \\
\hline

\end{tabular}
\label{tab:list_qr_code_features}
\end{table}

\subsubsection{Statistical Features}
These features are explicitly derived after the quantitative analysis and evaluation of black and white modules in QR codes. From Fig.~\ref{comparing_legit_phish_qr}, two QR codes (a) legitimate and (b) phishing of version 2 and 3 are presented. Both QR codes have 625 and 841 modules in total, respectively. Moreover, if we consider distribution of black and white modules separately then it is evident that legitimate code has 313 and 312, whereas phishing code has 440 and 441 black and white modules. This clearly indicates that both QR codes are different to each other in terms of modules representation. Similarly, we computed 19 statistical features of QR codes after quantitative evaluation including black-to-white ratio, density of QR code, mean density, row-wise and column-wise density, row-transitions and column-transitions, etc, which are also presented in Table. \ref{tab:list_qr_code_features}. After comparing the values of these features, it was apparent that legitimate and phishing QR codes have different patterns in the demonstration of modules which could be an aid in achieving the goal of classification of legitimate and phishing QR codes without extracting the embedded content. Here, it is worth mentioning that we are considering modules of QR codes not pixels. Therefore, features like density, entropy are computed on exact distribution of modules which is different from image processing techniques where whole QR code is taken as image and non-interpretable features are computed on the base of pixels not modules. Additionally, if the proposed approach is unable to found accurate arrangement of modules then all 19 statistical features may produce incorrect results which could also result in erroneous prediction outcomes. Given this, we ensured that our approach reliably produces precise output regarding quantitative analysis of QR code modules. Further details on how we observed the total modules in QR codes is explained in next section.

\section{QR\"{\i}S (QR: is it Safe?) The Proposed Methodology}
This section presents the proposed technique for Quishing classification based on the structural analysis of QR codes instead of extracting or analyzing their embedded content.\par

\begin{figure*}[htbp]
\begin{center}
\includegraphics[width=\linewidth]{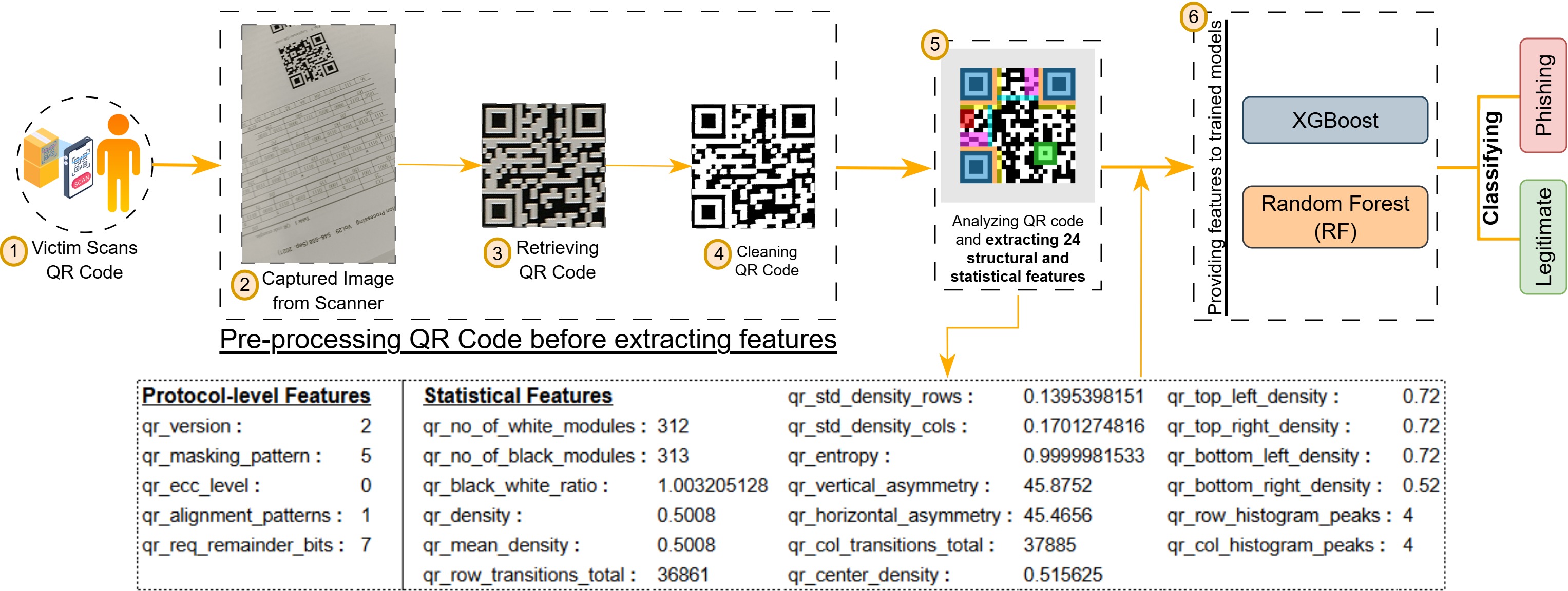}
\caption{A high-level architecture of QR\"{\i}S framework. This includes capturing the QR code image while user scanning the QR code. After copping the QR code part, further pre-processing steps are applied to make it clean before extracting structural features. An example of 24 extracted features is also presented, which later fed in to trained ML models (XGBoost and RF) to make the final prediction.}
\label{high_level_architecture}
\end{center}
\end{figure*}

\subsection{System Overview}
The overview of QR\"{\i}S framework for Quishing classification is shown in Fig.~\ref{high_level_architecture} which comprises of multiple sequence of steps. In step 1, a user scans a QR code using the built-in scanner from his/her smartphone device. Further in step 2, on detection of QR code, the QR\"{\i}S algorithm captures the QR code and its surrounding scene as an image in full with the help of device's camera. To retrieve only the QR part, we utilize a function \textit{QRCodeDetector} of python library `OpenCV' in step 3. The properties of this function not only extracts the QR code, but also aligns it to default alignment (which is required in our case). At this stage, QR code might be distorted or blurry because of the poor focus, lightning, or motion of cameras one uses. Therefore, in step 4, the algorithm further clean the QR code by applying various filters (i.e., median blur, gaussian blur, threshold, CLAHE) available in `OpenCV' library. The purpose of above pre-processing steps is to make sure that our approach should be able to efficiently recognize structural patterns of the QR code for the extraction of protocol-level and statistical features. With blurry or distorted image, the algorithm might not able to read correct modules for version and format bits which may further results in incorrect values of ECC level and masking pattern.\par
Additionally, for statistical features we are emphasized on exact number of modules present in the QR code instead of the pixels as discussed in Section III. Following this, in step 5, our approach analyzed the QR code and converted it into binary format (1,0) (details discussed in next subsection) to extract the structural features. To better comprehend this, Fig.~\ref{high_level_architecture} shows the output of 24 protocol-level and statistical features of a QR code extracted during our analysis. In step 6, these features are subsequently fed into the trained supervised ML models for prediction. It is to be noted that we trained two models, i.e., XGBoost and Random Forest (RF) on our datasets. Moreover, the stepwise operational flow of the proposed approach is also presented in Algorithm \ref{algo1}.\par
Note that the working of QR\"{\i}S approach in Fig.~\ref{high_level_architecture} does not need to extract and analyze the potentially harmful embedded content of QR codes thereby it strengthens users' privacy which is a key factor in protecting end-users. In fact, it explores the structural patterns of QR code image and classify that into correct cluster, i.e., legitimate or phishing. Accordingly, the proposed approach proves to be a preemptive and efficient to identify phishing threats delivered through QR codes which allows its integration with existing QR scanners without compromising user experience. Moreover, unlike existing solutions the utilization of lightweight ML models (XGBoost and Random Forest (RF)) makes it more useful for the resource-constrained environments like smartphones.\par

\subsection{Conversion of QR code into Binary Format}
\begin{figure}[htbp]
\begin{center}
\includegraphics[width=\linewidth]{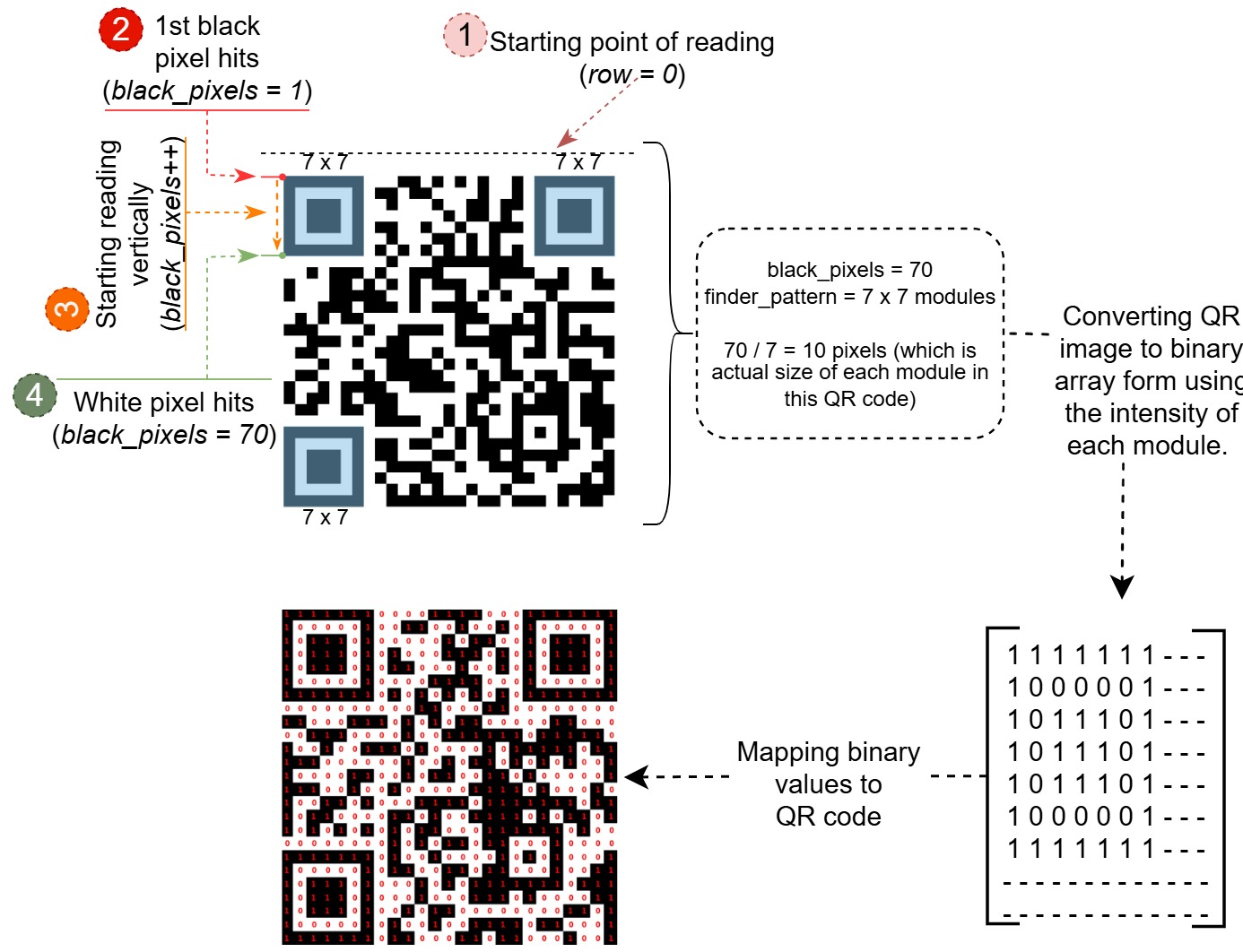}
\caption{A detailed overview of demonstrating the exact mapping conversion of QR code from black and white squares to binary format (1,0). This technique is can only work on QR codes which are composed of black and white squares.}
\label{qr_code_to_binary}
\end{center}
\end{figure}

The first step before extracting features from QR codes is to convert it into machine-readable format. It is important because to read the patterns of format information bits, locations of alignment patterns, and computing statistical features on the base of modules. For this, the algorithm transformed whole QR code into a binary grid array where black modules represented as 1 and white as 0. To accomplish this, we developed a systematic method presented in Fig.~\ref{qr_code_to_binary} and the relevant pseudocode is scripted in Algorithm \ref{algo1}. This method can work on any QR code that is build of black and white square modules. Firstly, the algorithm obtained the width of image in python using \textit{img.shape[:2]} (i.e., \textit{width = 250px} from Fig.~\ref{qr_code_to_binary} and Algorithm \ref{algo1}). Next, the algorithm estimated the size of each module in the QR code. For this, the algorithm selected a specific element of QR code which remains in same position and same structure in every QR code which is `Finder Patterns'. Afterwards, the algorithm performed the row wise traversal of QR code image matrix which scans each row from left to right until a black pixel is identified. This part represents the beginning of the QR code modules and position of top left finder pattern. From here, the algorithm started scanning vertically downwards until a white pixel hits which means the finder pattern region is ended and in parallel the algorithm keep counting the black pixels during this process (i.e., \textit{black\_pixels = 70} from Fig.~\ref{qr_code_to_binary}). Next, the algorithm divided the total number of black pixels [70] with 7 (as 7x7 modules is the fixed size of finder patterns). This gives us the exact size of module which is [10.00] as presented in Fig.~\ref{qr_code_to_binary}. Later on, the algorithm divided QR code image into grid according to the computed size of module and further calculated the average pixel intensity of each module to label them as 1 (black) or 0 (white). This process creates a binary array matrix which reflects the exact layout of QR image. To verify the obtained matrix, we draw it over the QR code in the form of grid lines (in red color) as shown in Fig.~\ref{qr_code_to_binary}.

\begin{algorithm}
\caption{Flow of QR\"{\i}S Method}
\label{algo1}
\begin{algorithmic}[1]
\scriptsize
\State \textbf{Require:} User's smartphone scanned and captured a QR code
\State $captured\_image \gets$ \Call{mobile\_scanner}{ }
\State $qr\_code \gets$ \Call{extract\_qr\_only}{captured\_image}
\State $cleaned\_qr\_code \gets$ \Call{apply\_filters}{qr\_code}
\State $binary\_form \gets$ \Call{convert\_qr\_into\_binary}{total\_modules}
\State $all\_features \gets$ \Call{extract\_all\_features}{binary\_form}
\State $prediction \gets$ \Call{trained\_model}{all\_features}
\State $print(prediction)$\;
\vspace{0.1cm}

\Function{convert\_qr\_into\_binary}{ }
    \State $qr\_width \gets$ cleaned\_qr\_code.shape[;2]
    \State $module\_size \gets$ \Call{estimate\_module\_size}{cleaned\_qr\_code}
    \State $total\_modules \gets$ \Call{round}{qr\_width / module\_size}
    \For{$i$ in \textbf{range}($total\_modules$)}
        \For{$j$ in \textbf{range}($total\_modules$)}
            \State $binary\_form \gets 1$ \textbf{if} $mean\_intensity(i,j) < 189$ \textbf{else} $0$
        \EndFor
    \EndFor
    \State \Return $binary\_form$
\EndFunction
\vspace{0.1cm}
\Function{estimate\_module\_size}{cleaned\_qr\_code}       % define function
    \State $qr\_pixel \gets \text{Threshold}(cleaned\_qr\_code,189,250)$
    \For{each $row$ with index $row\_index$ in $qr\_pixel$} \Comment{Row traversal}
        \For{each $pixel$ with index $col\_index$ in $row$}
            \State $bit \gets 1$ \textbf{if} $pixel == 0$ \textbf{else} $0$
            \If{$bit == 1$}
                \State $black\_count = 0$
                \For{$k$ in \textbf{range}($col\_index$, \textbf{len}($row$))}
                    \State $bit \gets 1$ \textbf{if} $row[k] == 0$ \textbf{else} $0$
                    \If{$bit == 1$} $black\_count++$
                    \Else $\ break$ \EndIf
                \EndFor $\ break$
            \EndIf
        \EndFor
        \If{$black\_count > 0$} $break$
        \EndIf
    \EndFor
    \State $module\_size = math.ceil(black\_count / 7)$
    \State \Return $module\_size$
\EndFunction

\end{algorithmic}
\end{algorithm}

\subsection{Model Training and Evaluation Metrics}
This section presents the settings we defined for training of the proposed model. Afterwards, we discussed various metrics selected for the evaluation of the proposed framework.

\subsubsection{Training Phase of QR\"{\i}S}
For training of the proposed framework, we selected two supervised ML models: XGBoost and Random Forest (RF). Both models trained on the extracted features of the QR codes discussed in Section IV-B. Furthermore, no feature transformation or categorical encoding was performed on the features because all of them fall in the numerical category which is also suitable for XGBoost and RF as shown in Fig.~\ref{high_level_architecture}. Overall, we have 200,000 balanced samples of QR codes features in each dataset (Feat-DataSet-1 and Feat-DataSet-2) which are further divided in 100000, 50000, 20000, 6000, 2000, 1000, and 200 balanced samples. We trained both models on both features datasets individually. We initiate our training with 200 samples and gradually increases the dataset size during training. Moreover, each of the dataset was randomly distributed into training 70\%, validation 15\%, and testing 15\% subsets using the stratified sampling technique. To fine-tune the performance of our models we used Optuna which is a well-known framework for hyperparameter tuning based on Bayesian optimization with pruning. Moreover, during this process we defined various hyperparameters for XGBoost and RF, which are presented in Table~\ref{tab:list_hyperparameters}. Optuna was set up to compute the mean accuracy as the objective function for the evaluation of performance using 5-fold stratified cross-validation on the training subset. Additionally, the study with Optuna was performed with 100 trials maximum or until a 1-hour timeout limit. Also, to decline the underperforming trials, the MediaPruner was utilized which activated after 5 warm-up steps and 10 start-up trials. With above settings, we ran 10 iterations for each model on both datasets and continuously monitor the effectiveness and consistency of the models.\par

As our proposed framework is designed to solve a binary classification problem, therefore, we considered various relevant evaluation metrics to assess the performance of QR\"{\i}S against both models. These metrics includes accuracy, precision, recall, F1-score, Area Under the Curve (AUC) score, and Receiver Operating Characteristic (ROC) curve.

\begin{table}[t]
\caption{Hyperparameters for XGBoost and Random Forest (RF)}
\centering
\renewcommand{\arraystretch}{1.2}
\begin{tabular}{>{\centering\arraybackslash}m{2.05cm}
                |>{\centering\arraybackslash}m{1.08cm}
                |>{\centering\arraybackslash}m{2.11cm}
                |>{\centering\arraybackslash}m{1.9cm}}
\hline
\textbf{Hyperparameter Name} & \textbf{XGBoost (Values)} & \textbf{Hyperparameter Name} & \textbf{Random Forest (RF) (Values)}\\
\hline
\hline

n\_estimators & 50, 300  & n\_estimators & 100, 1000 \\
\hline
max\_depth & 3, 15  & max\_depth & 5, 50 \\
\hline
learning\_rate & 0.01, 0.3  & min\_samples\_split & 2, 20 \\
\hline
subsample & 0.5, 1.0 & min\_samples\_leaf & 1, 20 \\
\hline
colsample\_bytree & 0.5, 1.0  & max\_features & 'sqrt', 'log2', None \\
\hline
gamma & 0, 5  & bootstrap & True, False \\
\hline
min\_child\_weight & 1, 10  & criterion & 'gini', 'entropy', 'log\_loss' \\
\hline
eval\_metric & 'logloss'  & - & - \\
\hline

\end{tabular}
\label{tab:list_hyperparameters}
\end{table}

\section{Experimental Results}
We performed our experiments on High Performance Computer (HPC) with python 3.12.8 also supporting 12 CPU cores, 32 GB of memory, and no GPU was involved during this training. The results from our experiments on datasets are presented in Table~\ref{tab:dataset_a_training_results}.\par

\begin{table*}[t]
\caption{The performance of both models on Feat-DataSet-1 gradually improves with the increase of samples size (also similar trend was observed for Feat-DataSet-2). For dataset Feat-DataSet-1, majority of the time among all samples XGBoost consistently performs better than RF in context of every evaluation metric. Specifically, for 200,000 samples XGBoost achieved a validation accuracy of 83.05\%, a test (unseen dataset) accuracy of 83.18\%, precision of 80.28\%, recall of 87.53\%, F1-score of 83.75\%, and AUC of 0.912. On the other hand, Random Forest shows a minor drop in performance and gain 82.85\% of validation accuracy, 82.89\% of test (unseen dataset) accuracy, 79.53\% of precision, 88.29\% of recall, 83.68\% of F1-score and 0.909 of AUC. These results indicates that XGBoost slightly achieved higher phishing detection rate while considering the structural features of QR codes as compared to legitimate ones. Similarly, for 200000 samples against dataset Feat-DataSet-2, XGBoost achieved a validation accuracy of 79.70\%, a test (unseen dataset) accuracy of 80.05\%, precision of 77.29\%, recall of 83.63\%, F1-score of 80.34\%, and AUC of 0.882. On the other hand, Random Forest shows almost equal performance and gain 79.74\% of validation accuracy, 80.21\% of test (unseen dataset) accuracy, 77.64\% of precision, 83.62\% of recall, 80.52\% of F1-score and 0.884 of AUC. These consistent results of XGBoost and RF across Feat-DataSet-2 reflects effective generalization ability of the models.}
\centering
\renewcommand{\arraystretch}{1.2}
\begin{tabular}{>{\centering\arraybackslash}m{1.9cm}
                |>{\centering\arraybackslash}m{1.3cm}
                |>{\centering\arraybackslash}m{1.3cm}
                |>{\centering\arraybackslash}m{1.4cm}
                |>{\centering\arraybackslash}m{1.4cm}
                |>{\centering\arraybackslash}m{1.8cm}
                |>{\centering\arraybackslash}m{1.6cm}
                |>{\centering\arraybackslash}m{1.7cm}
                |>{\centering\arraybackslash}m{1.4cm}}
\hline
\multirow{2}{*}{\parbox{1.5cm}{\centering \textbf{Features Dataset}}} & \multirow{2}{*}{\parbox{1.3cm}{\centering \textbf{Dataset Samples}}} & \multirow{2}{*}{{{\makecell{\textbf{Model}}}}} & \multicolumn{2}{c|}{\textbf{Accuracy (\%)}} & \multirow{2}{*}{{{\makecell{\textbf{Precision (\%)}}}}} & \multirow{2}{*}{{{\makecell{\textbf{Recall (\%)}}}}} & \multirow{2}{*}{{{\makecell{\textbf{F1-Score (\%)}}}}} & \multirow{2}{*}{{{\makecell{\textbf{AUC Score}}}}} \\
\cline{4-5}
& & & \textbf{Validation} & \textbf{Test} &  &  &  &  \\
\hline
\hline

\multirow{16}{*}{{{\makecell{Feat-DataSet-1}}}} & \multirow{2}{*}{{{\makecell{200}}}}  & XGBoost & 76.42 & 60.00 & 76.92 & 66.67 & 71.43 & 0.676 \\
&  & RF & 75.71 & 63.33 & 57.14 & 53.33 & 55.17 & 0.631 \\

\cline{2-9}
& \multirow{2}{*}{{{\makecell{1000}}}}  & XGBoost & 72.85 & 73.33 & 71.59 & 84.00 & 77.30 & 0.782 \\
&  & RF & 71.57 & 73.33 & 72.62 & 81.33 & 76.73 & 0.774 \\

\cline{2-9}
& \multirow{2}{*}{{{\makecell{2000}}}}  & XGBoost & 73.21 & 76.00 & 70.48 & 78.00 & 74.05 & 0.818 \\
& & RF & 71.92 & 77.00 & 72.35 & 82.00 & 76.88 & 0.841 \\

\cline{2-9}
& \multirow{2}{*}{{{\makecell{6000}}}}  & XGBoost & 75.71 & 78.11 & 74.90 & 81.56 & 78.09 & 0.849 \\
& & RF & 74.66 & 76.77 & 72.96 & 81.56 & 77.02 & 0.841 \\

\cline{2-9}
& \multirow{2}{*}{{{\makecell{20000}}}}  & XGBoost & 78.47 & 81.43 & 77.66 & 85.73 & 81.50 & 0.888 \\
&  & RF & 78.76 & 80.96 & 76.64 & 87.27 & 81.61 & 0.889 \\

\cline{2-9}
& \multirow{2}{*}{{{\makecell{50000}}}}  & XGBoost & 80.59 & 81.97 & 78.62 & 86.37 & 82.31 & 0.896 \\
&  & RF & 80.45 & 81.93 & 77.93 & 86.61 & 82.04 & 0.894 \\

\cline{2-9}
& \multirow{2}{*}{{{\makecell{100000}}}}  & XGBoost & 82.12 & 82.32 & 78.99 & 87.05 & 82.82 & 0.905 \\
&  & RF & 81.96 & 82.22 & 78.41 & 88.32 & 83.07 & 0.904 \\

\cline{2-9}
& \multirow{2}{*}{{{\makecell{200000}}}} & XGBoost & 83.05 & \textbf{83.18} & 80.28 & 87.53 & 83.75 & 0.912 \\
&  & RF & 82.85 & 82.89 & 79.53 & 88.29 & 83.68 & 0.909 \\

\hline
\multirow{2}{*}{{{\makecell{Feat-DataSet-2}}}} & \multirow{2}{*}{{{\makecell{200000}}}} & XGBoost & 79.70 & 80.05 & 77.29 & 83.63 & 80.34 & 0.882 \\
&  & RF & 79.74 & 80.21 & 77.64 & 83.62 & 80.52 & 0.884 \\
\hline

\end{tabular}
\label{tab:dataset_a_training_results}
\end{table*}

\begin{figure}[htbp]
\begin{center}
\includegraphics[width=\linewidth]{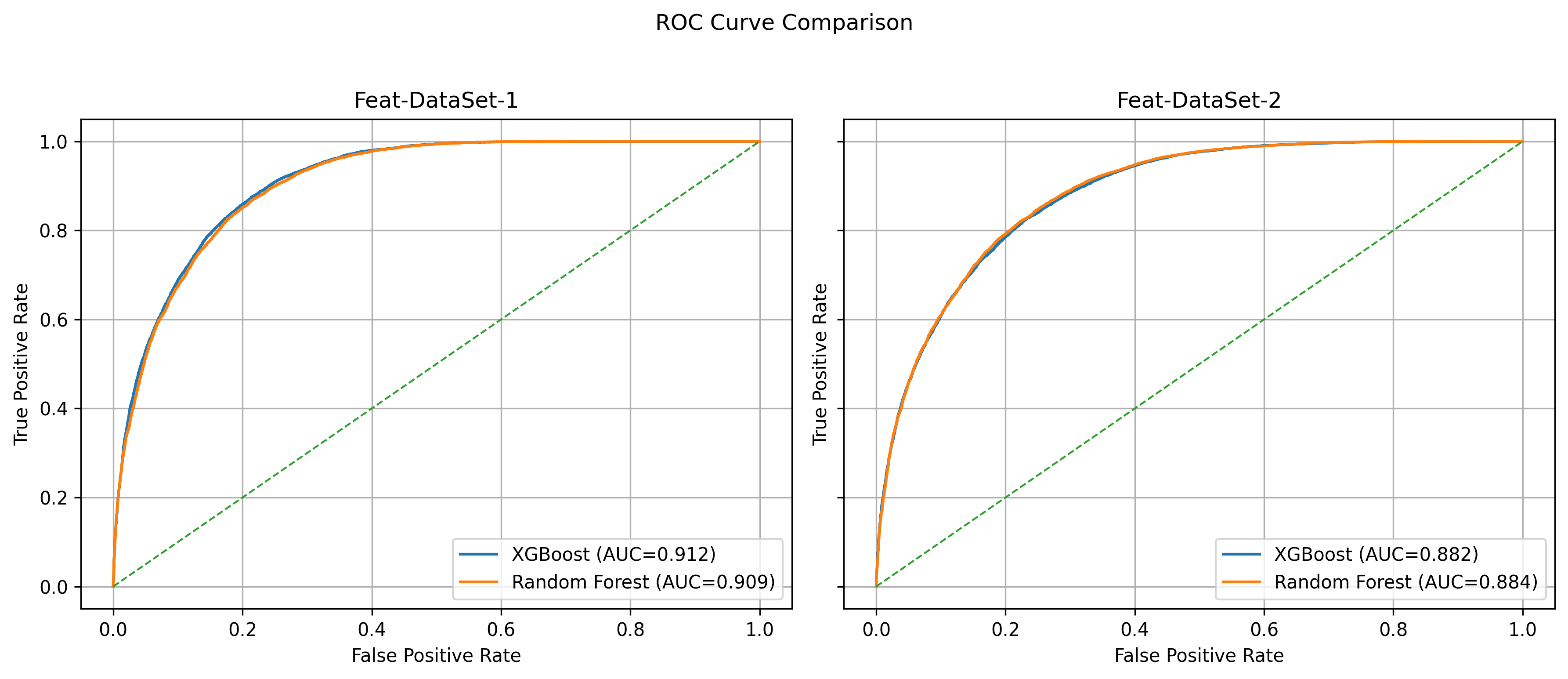}
\caption{For Feat-DataSet-1, both models maintains a high TPR rate which reflects that models were able to differentiate a wide range of phishing samples effectively while consistently keeping low FPR rate. The XGBoost curve remains a bit higher than RF curve from TPR value 0.3. This variation is due to the difference between the AUC values of both models. In case of Feat-DataSet-2, both curves almost coincide along the whole curve while RF slightly outperforming XGBoost at certain instances. Overall, the performance of both models across both datasets indicates their generalizabitiy and adaptability to learn the distinct patterns of legitimate and phishing QR codes on the base of structural features.}
\label{roc_comparison}
\end{center}
\end{figure}

Considering the overall comparison of models on both datasets, it is noteworthy that there is an average 3\% drop of performance of models across validation accuracy, test accuracy, precision, recall, and F1-score against Feat-DataSet-2. This might be the reason of structural variation in QR codes of Feat-DataSet-2 as compared to Feat-DataSet-1. For example, more diversity in QR code versions, ECC levels, density, or masking patterns, can potentially makes it challenging for the models to classify QR codes. Also, Fig.~\ref{roc_comparison} demonstrates the ROC curve of both models which is evaluated on both datasets and depicts the trade-off between sensitivity (True Positive Rate - TPR) and specificity (False Positive Rate - FPR).\par

\subsection{Interpretation of Our Results}
\begin{figure}[htbp]
\begin{center}
\includegraphics[width=\linewidth]{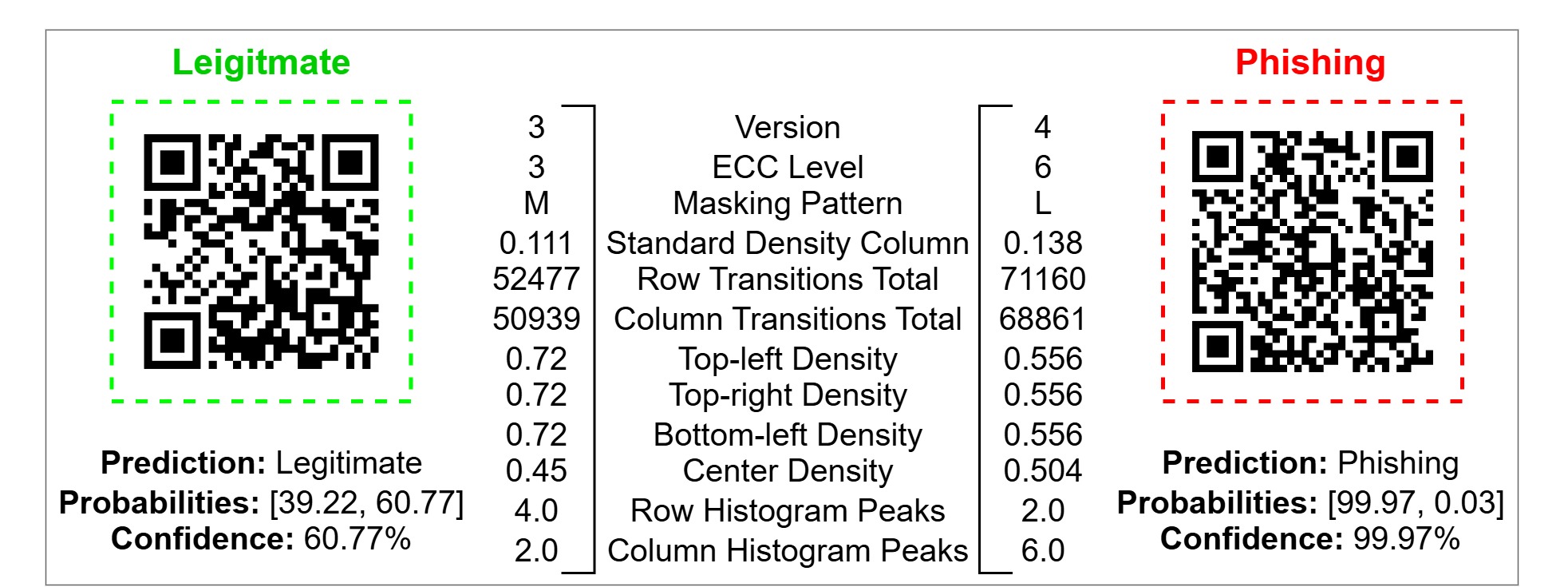}
\caption{A comparative example of legitimate and phishing QR codes from Feat-DataSet-1, showcasing the prediction results made by XGBoost model. Moreover, the actual interpretable features of both QR codes are presented to better comprehend the decision process of trained model using our approach.}
\label{interpretability_phish_legit}
\end{center}
\end{figure}

The proposed approach classified QR codes based on their extracted structural features. We emphasize that these features enable further understanding of the decision making process of ML models in QR code classification. Given this, the transparency is critical not only specifically for the evaluation of models' robustness but to enabling investigators to trust and understand the classification process. Regarding this, we provided an example case study in Fig.~\ref{interpretability_phish_legit} which illustrates the interpretability of two QR codes (one is legitimate and the other is phishing) and the outcome results are predicted by XGBoost. Also, in this specific scenario, the demonstrated values in Fig.~\ref{interpretability_phish_legit} are the dominated extracted features of both QR codes. Furthermore, it shows that the legitimate QR code has version 3 with high ECC level (M) ensures the strong error correction ability. Also, it highlights balanced quadrant densities (top-left, top-right, bottom-left), and consistent transition counts (row-wise and column-wise) possibly the indicators of proper and standard encoding of legitimate QR code. Henceforth, the model correctly predicted it as legitimate with confidence 60.77\%. In contrast, phishing QR code having higher version with low ECC level which allows more data capacity. Moreover, variations in quadrant density and relatively high values of transition counts, might reflecting the structural inconsistencies while encoding obfuscated or long phishing URLs and resultantly model predicted it as phishing with confidence 99.97\%. This high-level comparison represents that the predictions made by model are based on interpretable and measurable structural QR code features, therefore ensuring the credibility and trustworthiness of the proposed framework in real-world scenarios.

\vspace{-20pt}
\section{Comparative Analysis}
To the best of our knowledge, after reviewing the extensive literature we found that our proposed approach is not directly comparable to existing studies. This is because, we specifically extract the structural features of QR codes to perform Quishing classification which has never been explored before (in Quishing context). However, there are few studies~\cite{alaca2023cyber, minocha2024recognition, lourenco2023malicious} with which only a high level comparison can be possible due to the differences in their adapted methodology. For this, we have used accuracy, F1-score, and AUC score as our key metrics. At first, we have discussed the comparative studies~\cite{alaca2023cyber, minocha2024recognition, lourenco2023malicious, trad2025detecting} to help readers understand their presented work.\par
In~\cite{alaca2023cyber}, the authors proposed a hybrid model where they generated QR codes from the dataset CSE-CIC-IDS2018 which contains abnormal and unauthorised behavior of 6 different cyber attacks. Afterwards, they provided these QR codes to CNN models (MobilNetV2 and ShuffleNet) to extract 2000 in-depth features using Harris Hawk Optimization (HHO). With this approach, the highest accuracy they obtained was 95.89\% with the SVM model. However, their work lacks the explanation about creation of QR codes that how basic structural features are being assigned to each QR code including ECC level, version, etc. This is necessary to make sure that the embedded content is suitable according to the standardization of QR codes. Moreover, they employed black-box models to extract in-depth features which does not provide any interpretation of these features. Therefore, their approach lack clarity to identify that what specific QR codes features are dominating in classifying 6 different classes of cyber attacks which is also not comparable to our approach as we are particularly targeting Quishing attacks. Moreover, authors performed experiments on small dataset of 6000 samples (each attack category of 1000 samples) which further affects the generalizability of their approach in a broader context.\par
While the authors in~\cite{minocha2024recognition} utilized a dataset of 10,000 QR codes and used histogram density analysis to identify the features of QR codes. They implemented their approach with three CNN models ResNet50, MobileNetV3, and InceptionV2, from which InceptionV2 obtained the highest accuracy of 98.13\%. Their approach specifically dependent on analyzing the density aspect of provided limited dataset. Therefore, in case of any variation to the dataset may undermine the model's performance. Moreover, the features captured through histogram technique are non-interpretable which cannot be used to further clarify why the model decides that a specific QR code is legitimate or malicious.\par
Following this,~\cite{lourenco2023malicious} also proposed a deep learning technique using custom made CNN model, ResNet50, and VGG19 attaining accuracies of 97.13\%, 86.65\%, and 74.18\% respectively. The authors generated synthetic QR codes but with only 35,300 legitimate and 10,000 phishing URLs. However, no discussion is provided about any preliminary checks while embedding URLs in QR codes. Additionally, similar to~\cite{minocha2024recognition}, regardless of high accuracy their approach also lacks full ability to provide interpretability and transparency of the decision process.\par
The authors in~\cite{trad2025detecting} introduced a pixel-based ML technique to classify QR codes without looking into embedded content. They utilized a dataset of only 9987 QR code samples having identical properties including version, image size, box size, and border. In contrast, we created a novel dataset of 200,000 samples specifically based on the structural features of QR codes having diverse properties which makes our approach more adaptable to work with any QR code. Moreover, authors' reliance on pixel-based strategy is not capable to provide any structural level details as presented in this paper. Besides, few other studies~\cite{marappan2023enhancing, al2021secure} emphasized on classifying QR codes by extracting and analyzing the embedded content. From the comparative context, these studies are relied on retrieving various features related to QR code's content rather than QR code's structural features. Therefore, it is not possible to perform a comparison evaluation of our approach with them.\par

\begin{table}[t]
\caption{Comparative Analysis}
\centering
\renewcommand{\arraystretch}{1.2}
\begin{tabular}{>{\centering\arraybackslash}m{0.7cm}
                |>{\centering\arraybackslash}m{1.2cm}
                |>{\centering\arraybackslash}m{1.6cm}
                |>{\centering\arraybackslash}m{1.0cm}
                |>{\centering\arraybackslash}m{1.0cm}
                |>{\centering\arraybackslash}m{0.7cm}}
\hline
\multirow{3}{*}{{\makecell{\textbf{Ref.}}}} & \multirow{3}{*}{{\makecell{\textbf{Dataset}}}} & \multirow{3}{*}{{\makecell{\textbf{Models}}}} & \multicolumn{3}{>{\centering\arraybackslash}m{3.5cm}}{\textbf{Performance Metrics}} \\
\cline{4-6}
& & & \textbf{Acc. (\%)} & \textbf{F1-S. (\%)} & \textbf{AUC} \\
\hline
\hline

\multirow{2}{*}{{{\makecell{\textbf{Ours}}}}} & \multirow{2}{*}{{{\makecell{200,000*}}}} & XBoost & 83.18 & 83.75 & 0.912 \\
& & RF & 82.89 & 83.68 & 0.909 \\
\hline
\multirow{2}{*}{{{\makecell{\cite{alaca2023cyber}}}}} & \multirow{2}{*}{{{\makecell{6,000}}}} & SVM & 95.89 & 95.89 & - \\
& & kNN & 91.56 & 91.54 & - \\
\hline
\multirow{3}{*}{{{\makecell{\cite{minocha2024recognition}}}}} & \multirow{3}{*}{{{\makecell{10,000}}}} & MobileNetV3 & 96.87 & - & - \\
& & ResNet50 & 98.04 & - & - \\
& & InceptionV3 & 98.13 & - & - \\
\hline
\multirow{3}{*}{{{\makecell{\cite{lourenco2023malicious}}}}} & \multirow{3}{*}{{{\makecell{45,300}}}} & Custom-CNN & 97.13 & - & - \\
& & ResNet50 & 86.65 & - & - \\
& & VGG19 & 74.18 & - & - \\
\hline
\multirow{6}{*}{{{\makecell{\cite{trad2025detecting}}}}} & \multirow{6}{*}{{{\makecell{9,987}}}} & Logistic Regression & 79.83 & 78.67 & 0.873 \\
& & Decision Tree & 75.78 & 75.99 & 0.813 \\
& & RF & 79.93 & 77.46 & 0.892 \\
& & Na\"{\i}ve Bayes & 63.76 & 44.98 & 0.754 \\
& & XGBoost & 82.58 & 81.84 & 0.908 \\
& & LightGBM & 82.93 & 82.14 & 0.913 \\
\hline

\end{tabular}
\label{tab:comarative_analysis}
\vspace{1ex}
\parbox{1\linewidth}{\centering\footnotesize
\textbf{* :} \textit{Feat-DataSet-1}, \textbf{Acc. :} \textit{Accuracy}, \textbf{F1-S. :} \textit{F1-Score}}
\end{table}

From comparison with~\cite{alaca2023cyber, minocha2024recognition, lourenco2023malicious, trad2025detecting} in Table~\ref{tab:comarative_analysis}, we acknowledge that our models relatively achieved average accuracy than black-box model approaches. However, it is worth noting that none of the existing studies is capable to provide key details discussed in this paper for classification of QR codes. These approaches do not provide interpretability and transparency of internal decision process while distinguishing QR codes that leads to limited or no insights on why a QR code is detected as phishing or legitimate which are essentially required to understand and address security implications in context of Quishing. Other than that, the black-box techniques demands high computational cost~\cite{menghani2023efficient}, which makes them unsuitable for real-world deployment in resource-constrained environments (like smartphones).\par

\section{Mobile App Implementation}
\begin{figure}[htbp]
\begin{center}
\includegraphics[width=\linewidth]{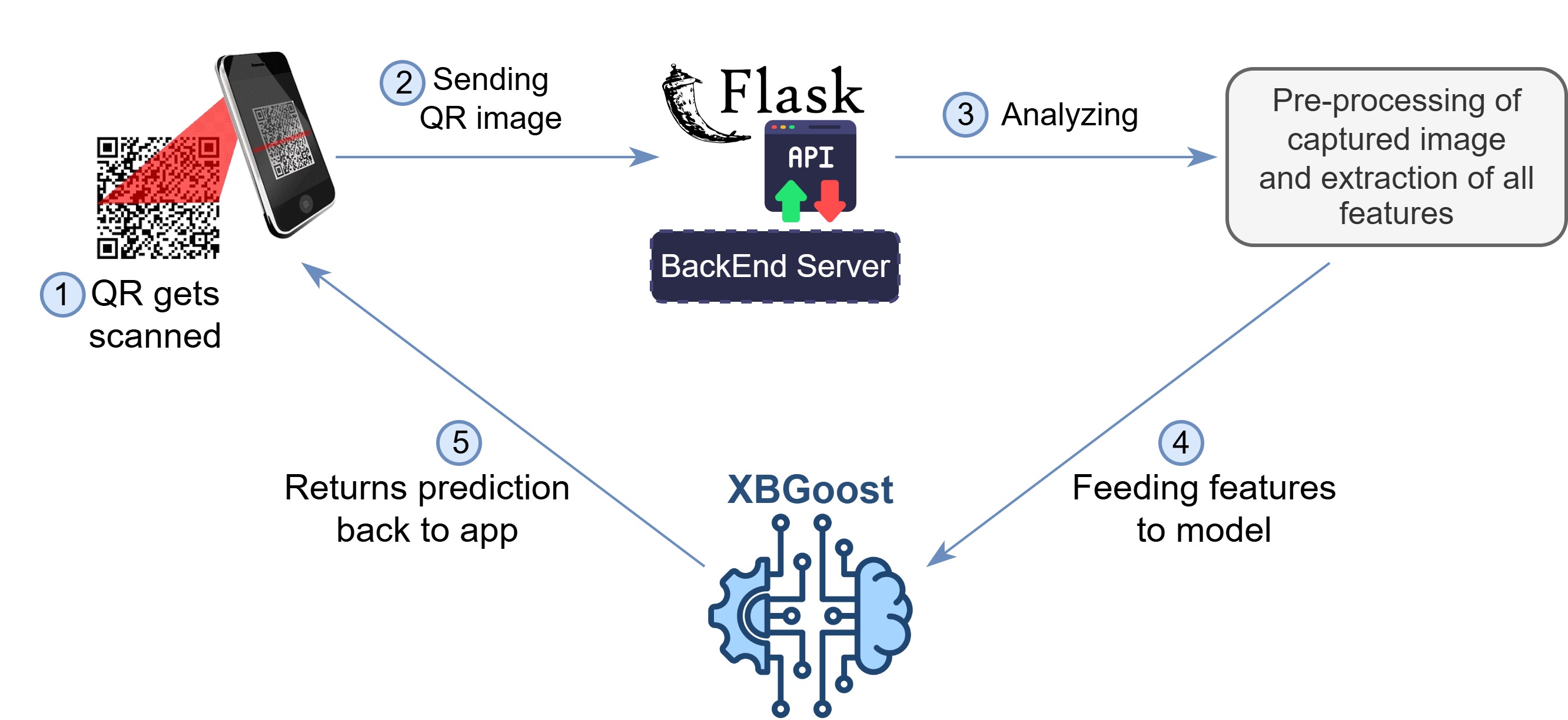}
\caption{Working of mobile app architecture which includes 5 major steps from scanning a real-world QR code to present the real-time detection results on user's mobile device.}
\label{mobile_app_architecture}
\end{center}
\end{figure}

For the validation of practical feasibility of {QR\"{\i}S} framework, we developed a mobile app (android) which can detect the legitimacy of QR codes in real-time. For development, we employed Flutter (which supports cross-platform i.e., Android and iOS). We utilized \textbf{\textit{(qr\_code\_scanner)}} plugin to detect QR code. Moreover, we used python Flask API as backend service to receive, process, and return the predicted results after analyzing QR code. The system architecture of mobile app is consists of 5 steps as presented in Fig.~\ref{mobile_app_architecture}. Firstly, an individual scans a QR code by keeping the smartphone in front of the QR code. On QR detection, it captures the image of QR code and forwards it to the Flask API in the second step. In step three, the backend script performs the pre-processing techniques on captured QR code as discussed in Section V-A. Furthermore, it extracts 24 protocol-level and statistical features of QR code and feed it to the trained model of XGBoost for prediction in step four. Afterwards, in step five, the predicted results sent back to the user screen on mobile app without extracting the embedded content of QR code.

\begin{figure}[htbp]
\begin{center}
\includegraphics[width=\linewidth]{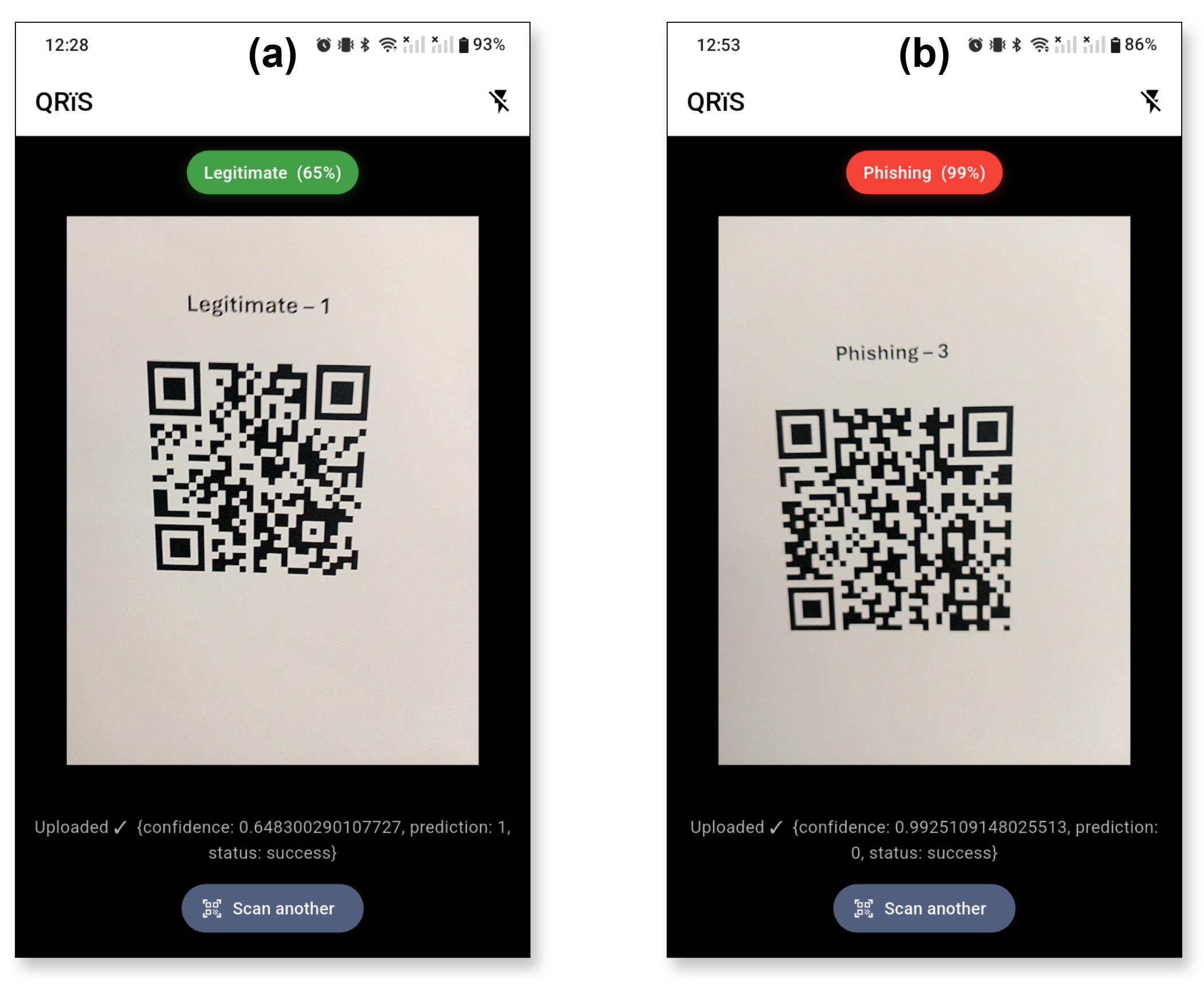}
\caption{Mobile app interface: (a) while scanning a legitimate QR code with the prediction as legitimate with 65\% confidence, (b) represents a phishing QR code with the prediction as phishing with 99\% confidence by the model.}
\label{app_results}
\end{center}
\end{figure}

For evaluation of mobile app, we randomly scanned 10 legitimate and 10 phishing QR codes from Feat-DataSet-2. The results are presented in Table~\ref{tab:mobile_app_results}. From 2 out of 10 legitimate QR codes, the app successfully detected them as legitimate whereas for phishing the successful outcome was 10 out of 10. For reference, we have presented interface of our mobile app in Fig.~\ref{app_results} (a) where it correctly detected a QR as legitimate and Fig.~\ref{app_results} (b) reflects the interface of phishing outcome. These promising results demonstrate the practical feasibility and potential of our proposed framework for real-world deployment where it can serve as a proactive defensive tool to mitigate Quishing attacks. Moreover, for research reproducibility, we have uploaded all the relevant source code of extracting protocol-level and statistical features, training of models, and mobile implementation to a publicly available Github repository\footnote{\url{https://github.com/mwahid905/curious-qr-code-phishing-solution}}.

\begin{table}[t]
\caption{Scan Results from Mobile App}
\centering
\renewcommand{\arraystretch}{1.2}
\begin{tabular}{>{\centering\arraybackslash}m{2.0cm}
                |>{\centering\arraybackslash}m{1.7cm}
                |>{\centering\arraybackslash}m{1.7cm}
                |>{\centering\arraybackslash}m{1.7cm}}
\hline
\textbf{QR Code Type} & \textbf{Total Scans} & \textbf{Successful} & \textbf{Unsuccessful} \\
\hline
\hline

Legitimate & 10  & 8 & 2 \\
Phishing & 10  & 10 & 0 \\
\hline

\end{tabular}
\label{tab:mobile_app_results}
\end{table}

\section{Discussions and Future Directions}
\begin{figure}[htbp]
\begin{center}
\includegraphics[width=\linewidth]{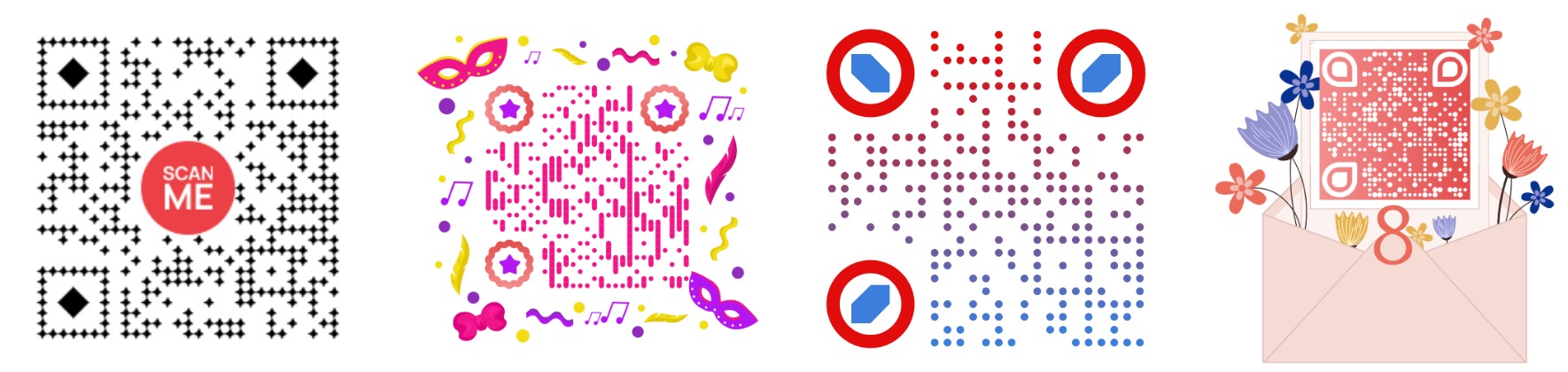}
\caption{An illustration of fancy QR Codes. Our proposed approach may not work on these QR codes because of the unique method of identifying module size and relevant features which is only possible with a QR code that is comprise of black and white square modules.}
\label{fancy_qr_codes}
\end{center}
\end{figure}

This paper presents a pioneer framework for early and before-hand classification of QR codes by extracting the structural features with no need to analyze QR code's embedded content. Nonetheless, our solution needs further improvements and we encourage both research and industry communities to contribute their valuable insights in addressing them. Firstly, {QR\"{\i}S} can specifically only works with QR codes which are composed of black and white squares as example provided in Fig.~\ref{qr_code_to_binary}. This is because of the technique we employed in converting QR code to binary format by estimating module size as discussed in Section V-B.
According to our technique, we perform a row wise traversal over the QR code from top left side until the first black module hits. To understand this, we have presented examples of QR codes with fancy layout of finder patterns and overall in Fig.~\ref{fancy_qr_codes}. This layout also includes utilization of gradient colors, revising module shapes, placing logo, and applying background colors, etc. In these particular scenarios, our technique needs further modifications to precisely identify module size which further limits the extraction of correct features of QR codes.\par

Secondly, the relatively low accuracy of both XGBoost and RF models against black-box model approaches need improvement. As we only extracted 24 protocol-level and statistical features of QR codes, whereas the consideration of other unknown relevant features may improve the accuracy. Apart from that, it is recommended to develop an ensemble learning approach which combines and works with both machine learning and deep learning techniques. By doing so, it may also contributes to enhance the overall performance of proposed detection framework.\par
Lastly, the current architecture of implemented mobile app is relied on internet connectivity because the backend server settings where the captured QR image is sent after scanning is running on python Flask API. To overcome this challenge, it will be more efficient to remove the API connection and perform the involved steps locally within app. In doing so, the app will function without relying on internet connection and individuals will be able to utilize this app to detect the legitimacy of QR codes particularly when there is no internet is available. Nonetheless, the presented framework is beneficial in many aspects, although it has few limitations which demands continued assessment of this early-stage work. Also in future, we aim to resolve the existing limitations of the proposed work.\par

\section{Conclusion}

This paper put forwards an innovative and pioneer method to classify the legitimacy of QR codes to mitigate Quishing attacks before-hand. We discussed that existing studies are either relied on extracting and analyzing embedded content of QR codes or based on black-box approaches which lack generalizability and demands high computation cost which also makes them unsuitable for real-world deployment. While we presented {QR\"{\i}S} which analyzes the structural patterns of QR codes. We extracted 24 protocol-level and statistical features from QR codes and prepared a features dataset of 400,000 blanched samples and shared it on publicly accessible Github repository along with relevant coding. We train the detection framework with two ML models: XGBoost and Random Forest (RF) attaining accuracies upto 83.18\% and 82.89\% respectively. Moreover, the results from our experiments highlights the effectiveness and practical feasibility of our approach in real-world scenarios. Furthermore, the comparison with existing approaches underscores the pressing need of the proposed framework. In future, we plan to extend our solution to ensure its universal applicability to detect any QR code including fancy ones.

% \section*{Acknowledgment}

%Dr. Reveryrand would like to acknowledge the funding by XLIM, Limoges, France. 

\ifCLASSOPTIONcaptionsoff
  \newpage
\fi


%\begin{thebibliography}{1}


\bibliographystyle{IEEEtran}
\bibliography{IEEEabrv,Bibliography}

%\end{thebibliography}
%\begin{biography}[{\includegraphics[width=1in,height=1.25in,clip,keepaspectratio]{mshell}}]{Michael Shell}

% ==== SWITCH OFF the BIO for submission
% ==== SWITCH OFF the BIO for submission
\begin{IEEEbiography}[{\includegraphics[width=1in,height=1.25in,clip,keepaspectratio]{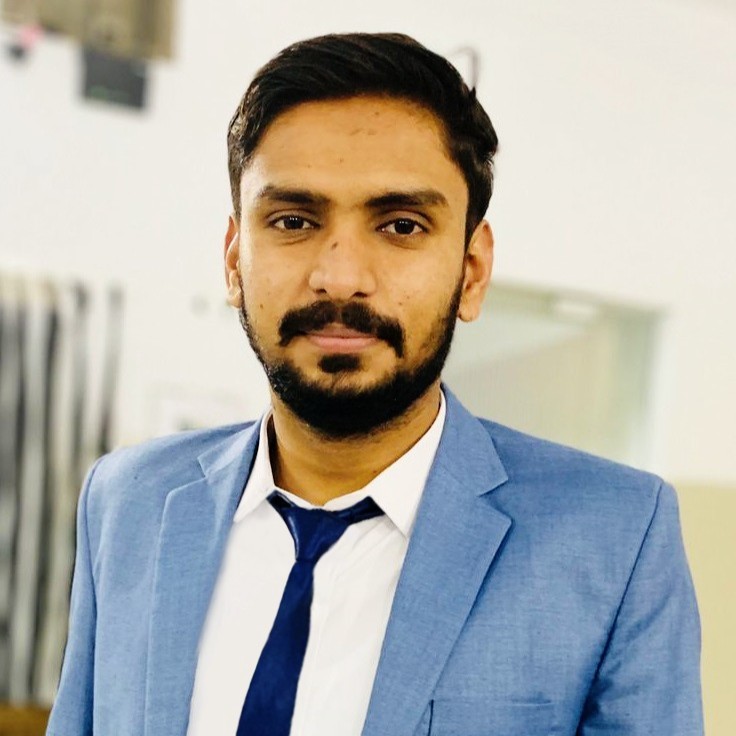}}]{Muhammad Wahid Akram} received his Bachelor degree (Hons.) and Master degree in Computer Science in 2018 and 2021, respectively. He is currently pursuing his PhD from Deakin University, Melbourne, Australia. His research interests includes cybersecurity, generative AI, machine learning, wireless communication, wireless networks, security and privacy.
\end{IEEEbiography}

\begin{IEEEbiography}
[{\includegraphics[width=1in,height=1.25in,clip,keepaspectratio]{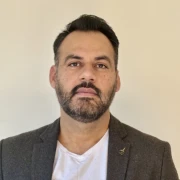}}]{Keshav Sood (Senior Member, IEEE)} received his B.Tech. degree (Hons.) in Electronic and Communication Engineering with distinction and the M.Tech. degree in Optical Fiber Engineering in 2007 and 2012, respectively. He was a trainee with the Terminal Ballistic Research Laboratory (TBRL, DRDO, Ministry of Defense) in Chandigarh, India. He received his Ph.D. degree in Information Technology (software-defined networking security) from Deakin University, Melbourne in 2018. He was the recipient of the Professor of IT award given by the School of IT for his outstanding academic achievements during his PhD degree. He completed his post-doctoral from The University of Newcastle, New South Wales, Australia. He is currently working as a Senior Lecturer in Cyber Security at Deakin University, School of IT. His work in cyber security for next generation networks has been published in top-notch security and networking venues.
\end{IEEEbiography}

\begin{IEEEbiography}
[{\includegraphics[width=1in,height=1.25in,clip,keepaspectratio]{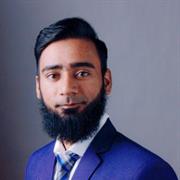}}]{Muneeb Ul Hassan (Member, IEEE)} is currently working as a Lectuer in School of Information Technology at Deakin University, Australia. He previously worked as Postdoctoral Researcher in Security \& Privacy at Swinburne University of Technology, Australia. did his Ph.D Degree from Swinburne University of Technology, Australia in 2021. He has been awarded Fellowship of Advance HE, UK in 2025. He is also a recipient of IEEE TCSC 2022 Award for Excellence (Early Career Researcher) in Scalable Computing for research excellence in privacy preservation of blockchain and decentralized energy systems. Alongside this, he has also won several Top Peer Reviewer Awards from Clarivate - Web of Science. He has published his research works in Top Tier Journals of the field including IEEE Transactions on Knowledge Data Engineering, IEEE Transactions on Services Computing, IEEE Communications Surveys \& Tutorials, etc. His main research interests include privacy protection, differential privacy, blockchain technology, cybersecurity, smart grid, cognitive radio ad hoc networks, artificial intelligence, cloud computing, big data, security and privacy, wireless networks, cognitive radio sensor networks, and mobile ad-hoc networks.
\end{IEEEbiography}

\vfill

\end{document}